\pdfoutput=1
 
\newcommand*{\ATLASLATEXPATH}{}
\documentclass[txfonts=true, texlive=2017, cernpreprint, USenglish]{\ATLASLATEXPATH atlasdoc}
 
\usepackage{\ATLASLATEXPATH atlaspackage}
\usepackage{\ATLASLATEXPATH atlasbiblatex}
\usepackage{\ATLASLATEXPATH atlascontribute}
\usepackage{\ATLASLATEXPATH atlasphysics}
 
\graphicspath{{logos/}{figures/}}

\usepackage{ANA-HION-2019-03-PAPER-defs}
\usepackage{multirow}
\usepackage{caption}
\usepackage{float}
\AtlasTitle{
Measurement of the azimuthal anisotropy of charged-particle
production in Xe+Xe collisions at $\sqrt{s_{\mathrm{NN}}}=5.44$~TeV
with the ATLAS detector
}

\date{\today}
 
\AtlasJournal{Phys.\ Rev.\ C.}
\PreprintIdNumber{CERN-EP-2019-227}
 
\AtlasAbstract{
This paper describes the measurements of flow harmonics $v_2$--$v_6$ in
3~$\mu\mathrm{b}^{-1}$ of Xe+Xe collisions at $\sqrt{s_{\mathrm{NN}}}=5.44$~TeV
performed using the ATLAS detector at the LHC.
Measurements of the centrality, multiplicity and $p_{\mathrm{T}}$ dependence
of the $v_n$ obtained using two-particle correlations and
the scalar product technique are presented.
The measurements are also performed using a template-fit procedure,
which was developed to remove non-flow correlations in small
collision systems.
This non-flow removal is shown to have a significant influence on the
measured $v_n$ at high $p_{\mathrm{T}}$, especially in peripheral events.
Comparisons of the measured $v_n$ with measurements in Pb+Pb
collisions and $p$+Pb collisions at $\sqrt{s_{\mathrm{NN}}}=5.02$~TeV
are also presented.
The $v_n$ values in Xe+Xe collisions are observed to be larger than
those in Pb+Pb collisions for $n=2$, 3 and 4 in the most central events.
However, with decreasing centrality or increasing harmonic order $n$, the
$v_n$ values in Xe+Xe collisions become smaller than those in Pb+Pb collisions.
The $v_n$ in Xe+Xe and Pb+Pb collisions are also compared as a function of the
mean number of participating nucleons, $\langle N_\text{part} \rangle$,
and the measured charged-particle multiplicity in the detector.
The $v_3$ values in Xe+Xe and Pb+Pb collisions are observed to be similar at
the same $\langle N_\text{part}\rangle$ or multiplicity,
but the other harmonics are significantly different.
The ratios of the measured $v_n$ in Xe+Xe and Pb+Pb collisions, as a function of centrality,
are also compared to theoretical calculations.
}
 
\AtlasJournalRef{Phys. Rev. C 101 (2020) 024906}
\AtlasDOI{10.1103/PhysRevC.101.024906}

\hypersetup{pdftitle={ATLAS document},pdfauthor={The ATLAS Collaboration}}
 
\begin{document}
 
\maketitle

\section{Introduction}
Heavy-ion collisions, such as those at the Relativistic Heavy Ion
Collider (RHIC)~\cite{Adams:2005dq,Back:2004je,Adcox:2004mh,Arsene:2004fa}
and at the Large Hadron Collider (LHC)~\cite{HION-2011-05,HION-2016-06,PhysRevLett.105.252302,CMS-HIN-10-002,Acharya:2018ihu,CMS-HIN-18-001},
produce a novel state of matter, with deconfined quarks and gluons,
commonly called quark--gluon plasma (QGP).
The QGP produced in such collisions expands anisotropically due to
spatial anisotropies in the initial geometry, which produce asymmetric
pressure gradients between the medium and the outside vacuum.
The anisotropic expansion leads to large azimuthal modulations in the
final distributions of the produced particles~\cite{PhysRevD.46.229}.
The single-particle azimuthal yields of particles produced in heavy-ion
collisions are typically characterized as a Fourier
series~\cite{Voloshin:1994mz} :
\begin{eqnarray}
\frac{dN}{d\phi}=\frac{N_{0}}{2\pi}\left(1+2\Sigma_{n=1}^{\infty}v_{n}\cos(n(\phi-\Phi_{n}))\right),
\label{eq:single}
\end{eqnarray}
where $\phi$ is the azimuthal angle of the particle momentum, and
the $v_{n}$ and $\Phi_{n}$ are the magnitude and
phase of the $n^{\mathrm{th}}$-order anisotropy.
The \vn\ are referred to as flow harmonics, while the $\Phi_{n}$
are referred to as event-plane angles.
The \vn\ are functions of the transverse momentum (\pT), pseudorapidity\footnote{\AtlasCoordFootnote} ($\eta$),
event-multiplicity and particle species, and fluctuate from event to event.
 
Measurements of the \vn\ and their comparisons with calculations based on
relativistic hydrodynamics have shown that the QGP produced in heavy-ion
collisions behaves like a nearly perfect fluid, characterized by a very low
ratio of shear-viscosity to entropy density, $\eta/s$,
close to the conjectured lower limit of $\hbar/4\pi k_{\mathrm{B}}$~\cite{Kovtun:2004de}.
Significant experimental progress has been made in recent years in precision
measurements of the \vn~\cite{HION-2011-01,HION-2011-05,HION-2016-06,PhysRevLett.105.252302,CMS-HIN-10-002,CMS-HIN-15-014},
event-by-event fluctuations in the \vn~\cite{HION-2012-10,HION-2012-04,CMS-HIN-16-019}
and the correlations between the magnitudes~\cite{HION-2014-03,ALICE:2016kpq} and
phases~\cite{HION-2012-03} of different-order anisotropies.
These high-precision measurements have led to significant improvement
in constraining the value of $\eta/s$~\cite{Gale:2013da,Heinz:2013th}.
However, due to large theoretical uncertainties in the understanding of the
initial stages of heavy-ion collisions, the precise value of $\eta/s$ or its
exact temperature dependence still remains unknown.
A possible data-driven way of further constraining the value of $\eta/s$,
and at the same time improving the understanding of the initial entropy
production in heavy-ion collisions, is to compare the \vn\ measured across
collision systems of different sizes~\cite{Giacalone:2017dud}.

This \papertype\ presents \vn\ measurements in \XeXe\ collisions at
a center-of-mass energy per nucleon pair of $\snn=5.44$~\TeV\ by ATLAS
using an integrated luminosity of 3~$\mu\mathrm{b}^{-1}$.
The flow measurements are performed with the two-particle correlation (\tpc)
and scalar product (SP) methods.
The measurements are also performed using a template-fit
procedure~\cite{HION-2015-09,HION-2016-01} developed by ATLAS to
measure correlations in small systems -- such as \pp\ and \pPb.
The template-fit procedure removes `non-flow' correlations that arise
from back-to-back jet pairs (dijets) and particle decays, which typically bias
the \vn\ measurements in low-multiplicity events, especially at high \pt.
 
The Xe nucleus is small compared to the Pb nucleus
-- the specific isotopes used at the LHC are $^{129}\mathrm{Xe}_{54}$ and $^{208}\mathrm{Pb}_{82}$.
Thus \XeXe\ collisions are expected to have
larger event-by-event fluctuations in the initial geometry compared
to \PbPb\ collisions~\cite{Giacalone:2017dud}.
On the other hand, a smaller system implies larger viscous effects in
the hydrodynamic expansion of the produced QGP
fireball~\cite{Giacalone:2017dud,Baier:2007ix,Romatschke:2007mq}.
Thus the \vn\ measurements in \XeXe\ collisions, and their comparison with those in \PbPb\
collisions allow the interplay of these two effects to be studied.

The outline of the \papertype\ is as follows:
Section~\ref{sec:experiment} describes the ATLAS detector subsystems
used in this measurement.
Section~\ref{sec:datasets} describes the dataset, the event selection
and requirements on charged-particle tracks used in the analysis.
Section~\ref{sec:methodology} gives a brief description of the \tpc,
template-fit and SP methods.
The systematic uncertainties in the \vn\ measurements are described
in Section~\ref{sec:systematics}.
In Section~\ref{sec:results} the results of the measurements
are presented, which include the \pt, centrality, and multiplicity
dependence of the \vn\, as well as its dependence on the mean number
of participating nucleons in the collisions ($\langle\npart\rangle$).
In Section~\ref{sec:comparison}, the present \XeXe\ \vn\ measurements
are compared with previous \vn\ measurements in \PbPb\ and
\pPb\ collisions.
Finally, Section~\ref{sec:summary} summarizes the results.

\section{The ATLAS detector\label{sec:experiment}}
The measurements presented in this paper were performed using the ATLAS detector~\cite{PERF-2007-01} at the LHC.
The principal components used in this analysis are the
inner detector~(ID),
calorimeter,
and the trigger and data acquisition systems.
 
The ID consisting of a silicon pixel detector, a silicon microstrip tracker
and a transition radiation tracker is immersed in a \SI{2}{\tesla}
axial magnetic field.
The ID provides charged-particle tracking in the range $|\eta| < 2.5$.
The high-granularity silicon pixel detector covers the interaction region and
typically provides four measurements per track. The first hit is
normally in the  `insertable B-layer' (IBL)~\cite{ATLAS-TDR-2010-19,PIX-2018-001},
which was installed before the 2015 data taking period.
The pixel detector is followed by the silicon microstrip tracker (SCT), which
typically provides measurements of four three-dimensional points per track.
These silicon detectors are complemented by the transition radiation
tracker, which enables radially extended track reconstruction up to
$|\eta| = 2.0$, providing around 30 hits per track.
The ATLAS calorimeter system consists of a liquid argon (LAr) electromagnetic
calorimeter covering $|\eta|<3.2$, a steel--scintillator sampling hadronic
calorimeter covering $|\eta|<1.7$, a~LAr hadronic calorimeter covering
$1.5<|\eta|<3.2$, and two LAr electromagnetic and hadronic forward calorimeters
(FCal) covering $3.2<|\eta|<4.9$.
The ATLAS trigger system~\cite{TRIG-2016-01} consists of a first-level (L1)
trigger implemented using a combination of dedicated electronics and
programmable logic, and a software-based high-level trigger (HLT).

\section{Dataset, event and track selections~\label{sec:datasets}}
The Xe+Xe data used in this \papertype\ were collected in October 2017.
A set of minimum-bias events was selected by a pair of complementary and
mutually exclusive triggers.
The first trigger required the total transverse energy deposited
in the calorimeters at L1 ($E_{\mathrm{T}}^{\mathrm{L1}}$)
to be larger than 4~\GeV, without any additional requirement at the HLT.
The second trigger required that
$E_{\mathrm{T}}^{\mathrm{L1}}$ be less than 4~\GeV\
with the additional requirement of a reconstructed track with $\pt>0.2$~\GeV\ at the HLT.
Together these two triggers selected all events with either $E_{\mathrm{T}}^{\mathrm{L1}}>4~\GeV$
or a reconstructed track at the HLT.
 
\begin{figure}[!htb]
\centering
\includegraphics[width=0.6\linewidth]{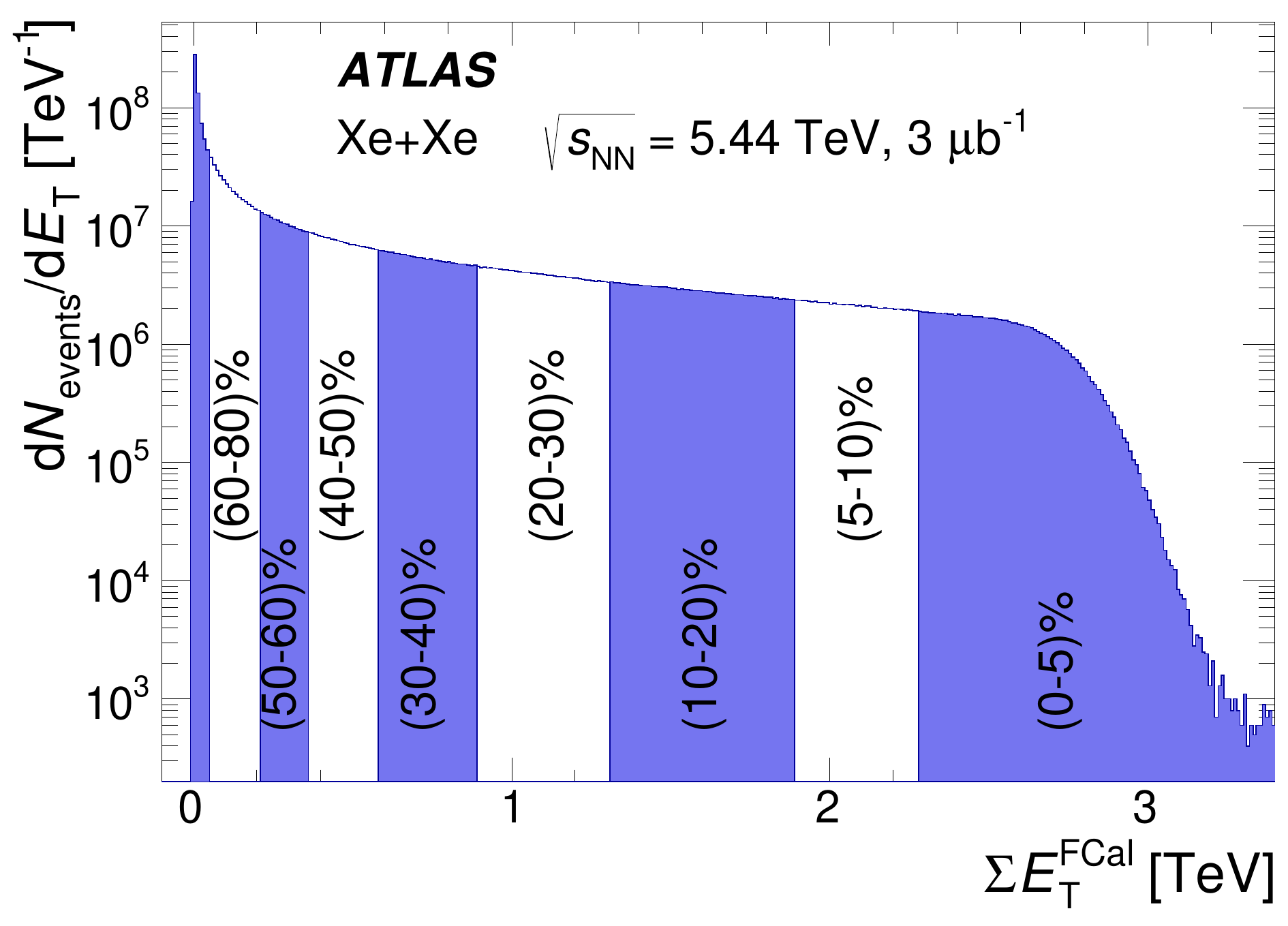}
\caption{
The $\Sigma E_{\text{T}}^{\text{FCal}}$ distribution in minimum-bias events together
with the selections used to define centrality classes covering the 0--80\% centrality range.
}
\label{fig:centrality}
\end{figure}

In the offline analysis, the $z$-coordinate of the primary vertex~\cite{ATL-PHYS-PUB-2015-026} is
required to be within 10~cm of the nominal interaction point.
Events containing more than one inelastic interaction (pileup events)
were removed by exploiting the correlation between
the transverse energy measured in the FCal (\fcalsum) and the number of tracks
associated with the primary vertex.
In a typical pileup event the track multiplicity associated with the primary
vertex belongs to a single \XeXe\ collision, while the energy deposited in calorimeters
contains contributions from multiple collisions.
Therefore, events with small values of measured multiplicity
and large \fcalsum, which differ markedly from those of the majority of \XeXe\
collisions, are removed from the analysis~\cite{HION-2012-04}.
The fraction of pileup events is estimated to be $\sim$0.1\%.
As in previous ATLAS heavy-ion analyses, the events are classified into
centrality percentiles based on the total transverse energy deposited
in the FCal in the event~\cite{HION-2011-01,HION-2011-05}.
The Glauber model~\cite{Miller:2007ri} is used to obtain a
correspondence between the \fcalsum\ distribution and the sampling
fraction of the total inelastic Xe+Xe cross-section, allowing the
setting of the centrality percentiles~\cite{HION-2011-01,HION-2011-05}.
The Glauber model is also used to obtain the mapping from the observed
\fcalsum to the primary properties, such as the mean number of nucleons
participating in the nuclear collision, $\langle\npart\rangle$, for
each centrality interval.
Figure~\ref{fig:centrality} shows the distribution of \fcalsum\ in data
and thresholds for the selection of several centrality intervals.
For the centrality dependence study, this analysis is restricted to
the \cent{0--80} most central collisions where the triggers are
fully efficient.

Charged-particle tracks are reconstructed from the signals in the ID.
A reconstruction procedure developed for tracking in dense environments
in \pp\ collisions, and optimized for heavy-ion collisions, is used
for this purpose~\cite{ATL-PHYS-PUB-2015-006}.
In the analysis, the set of reconstructed tracks is filtered using
several selection criteria.
For the nominal selection, the tracks are required to have
$\pT>0.5~\GeV$,
$|\eta|<2.5$,
at least two pixel hits,
with the additional requirement of a hit in the IBL when one is expected,
at least eight SCT hits,
and no missing hits in the pixel or SCT.
A hit is expected if the extrapolated track crosses an active
region of a silicon-sensor module (pixel or SCT) that has not been disabled, and a hit is said
to be `missing' when it is expected but not found.
Further, the transverse ($d_0$) and longitudinal ($z_0\sin(\theta)$)
impact parameters of the track relative to the primary vertex are required
to be less than 1~mm.
The track-fit quality parameter $\chi^2/$ndof is required
to be less than 6.
For cross-checks and for estimating systematic uncertainties
a looser and a tighter set of requirements are used.
For the looser selection, referred to as `loose', the
requirements on the number of pixel and SCT hits are
reduced to one and six, respectively, and the requirement on
$d_0$ and $z_0\sin\theta$ are relaxed to 1.5~mm.
These looser requirements on the tracks are also used
to study the multiplicity dependence of the \vn\
as they enhance the per-event multiplicity.
For the tighter selection, referred to as `tight', the topological requirements on
the reconstructed track are not altered, but the transverse and longitudinal
impact parameters of the track are restricted to be less than 0.5~mm.

In order to study the performance of the ATLAS detector,
a sample of 1M minimum-bias Xe+Xe Monte Carlo (MC) events was generated
using  HIJING version 1.38b~\cite{Wang:1991}.
Since the HIJING event-generator does not have any intrinsic mechanism to generate flow,
the latter is added after the generation using an
`afterburner' procedure~\cite{Jia:2013tja}, which slightly shifts the $\phi$
positions of generated particles to mimic flow.
The generated sample was passed through a full simulation of the ATLAS
detector~\cite{SOFT-2010-01} using \GEANT\!4~\cite{Agostinelli:2002hh}, and the MC events
were reconstructed by the same algorithms as the data.
The reconstructed particles in the MC events are used to calculate the
reconstruction efficiency -- the fraction of the generated charged
particles that are successfully reconstructed -- as a function of \pt\ and $\eta$,
and denoted by $\epsilon(\pt,\eta)$ below.
At mid-rapidity ($|\eta|<1$) and for events in the 0--5\% centrality interval,
the reconstruction efficiency is $\sim$60\%
at low \pt\ and increases to $\sim$73\% at higher \pt.
For $|\eta|>1$ the efficiency decreases to about 40--60\% depending on the \pt and centrality.
The reconstruction efficiency depends weakly on the centrality for low-\pt\ tracks,
for which it is smaller in the most central events by about 3\%
as compared to peripheral events.
For tracks with $\pt>1~\GeV$ the dependence on centrality is less than 1\%.
The MC simulation is also used to calculate the fraction of fake tracks
-- the fraction of the reconstructed particles that do not
correspond to any generated particle -- also as a function of
\pt\ and $\eta$, and denoted by $f(\pt,\eta)$ below.
The fake rate is less than $1\%$ across the \pt\ and centrality range used
in this analysis.
Additionally, systematic differences seen between the \vn\
evaluated using the reconstructed particles in the MC events -- using the
same techniques as in the data analysis -- and the
\vn\ implemented for the generated particles, are used
as multiplicative correction factors in the data analysis.
The size of this correction is discussed in Section~\ref{sec:systematics}.

The \XeXe\ measurements in this \papertype\ are compared with
similar measurements in \PbPb\ and \pPb\ collisions at $\snn=5.02$~\TeV.
Further, for the template fits, data from \pp\ collisions
at $\sqs=5.02$~\TeV\ are used~\cite{HION-2016-01} (see Section~\ref{sec:template}).
The \PbPb\ dataset, triggers and track selections used for the comparisons are
identical to those used in Ref.~\cite{HION-2016-06}.
The \pp\ and \pPb\ datasets, triggers and track selections are
identical to those used in Ref.~\cite{HION-2016-01}.
 
\section{Methodology\label{sec:methodology}}
In this section, a brief description of the methods used to measure
flow harmonics is provided.
First, the two-particle correlation analysis, including the construction of correlation
functions, is described.
Next, the template-fit procedure is explained, which starts from the \tpcs\ but implements
additional non-flow subtraction.
Finally, the steps involved in obtaining the SP results are presented.

\subsection{Two-particle correlations\label{sec:2pc}}
The \tpc\ method has been used extensively for flow measurements at
RHIC and the LHC~\cite{Adare:2008ae,HION-2011-01,HION-2012-10,HION-2012-13,
HION-2013-04,HION-2014-03,HION-2015-09,HION-2016-01,
HION-2016-06}.
In the \tpc\ method, the distribution of particle pairs
in relative azimuthal angle $\dphi=\phia-\phib$ and pseudorapidity
separation $\deta=\etaa-\etab$ is measured.
The particles $a$ and $b$ are conventionally referred to as the
`reference' and `associated' particles, respectively.
In this analysis, the two particles are charged particles measured by
the ATLAS tracking system, over the full azimuth and $|\eta|<2.5$,
resulting in a pair-acceptance coverage of $\pm5.0$
units in \deta.
In order to account for the detector acceptance effects, such
as localized regions with lower efficiency or dead regions,
the correlation is constructed from the ratio of the
distribution in which the reference and associated particles
are taken from the same event to the distribution in which
the reference and associated particles are taken from
two different events.
The correlation function is defined as:
\begin{eqnarray*}
\ctwo =\frac{S(\deta,\dphi)}{B(\deta,\dphi)}\, ,
\end{eqnarray*}
where $S$ and $B$ represent the `same-event' and `mixed-event'
pair distributions, respectively~\cite{HION-2011-01}.
The same-event distribution includes both the physical correlations
and the correlations arising from detector acceptance effects.
On the other hand, the mixed-event distribution reflects only the
effects of detector inefficiencies and nonuniformity, but contains
no physical correlations.
Detector acceptance effects largely cancel out in the $S/B$ ratio~\cite{Adare:2008ae}.
To ensure that the acceptance effects in the $B$ distribution closely match those
in the $S$ distribution, the $B$ distribution is constructed from
particles from two different events that have similar centrality
(or multiplicity) and $z$-vertex.
When constructing $S$ and $B$, in order to account for the inefficiency
in track reconstruction and for misreconstructed tracks,
pairs are weighted by their fake-track rates and the inverse product of
their reconstruction efficiencies (see Section~\ref{sec:datasets}):
\begin{eqnarray*}
(1-f(\pta,\etaa))(1-f(\ptb,\etab))/(\epsilon(\pta,\etaa)\epsilon(\ptb,\etab)).
\end{eqnarray*}
 
To investigate the \dphi-dependence of the long-range ($|\deta|>2$)
correlation  in more detail, the $S$ and $B$ distributions are integrated over
$2<\deta<5$ and a one-dimensional correlation function \ctwophi\
is constructed as follows:
\begin{eqnarray*}
\ctwophi = \frac{\int_{2}^{5} d|\deta| \; S(|\deta|,\dphi)}{\int_{2}^{5}
d|\deta| \; B(|\deta|,\dphi)}\equiv\frac{S(\dphi)}{B(\dphi)}.
\end{eqnarray*}
The $|\deta|>2$ requirement is imposed to reject the short-range
correlations and focus on the
long-range features of the correlation functions.
In a similar fashion to the single-particle distribution
(Eq.~\eqref{eq:single}), the \ctwophi\ can be expanded as a
Fourier series~\cite{HION-2011-01}:
\begin{eqnarray}
\ctwophi=C_{0}\left(1+2\Sigma_{n=1}^{\infty}\vnn(\pta,\ptb)\cos(n\dphi)\right).
\label{eq:2pc_fourier}
\end{eqnarray}
In the absence of any non-flow correlations, it can be shown that the
Fourier coefficients of the \ctwophi\ factorize into the product of
single-particle anisotropies as~\cite{Adare:2008ae}:
\begin{eqnarray}
\vnn(\pta,\ptb)=\vn(\pta)\vn(\ptb).
\label{eq:factorization}
\end{eqnarray}
The factorization of \vnn\ given by Eq.~\eqref{eq:factorization}
is expected to break down at high \pt\ and in low-multiplicity events,
where the \vnn\ measurements
are biased by non-flow correlations~\cite{HION-2011-01,HION-2016-06}.
The factorization is also expected to break down when the $\eta$
separation between the particles is small, and short-range
correlations dominate~\cite{HION-2011-01}.
However, the $|\deta|>2$ requirement removes most of such short-range
correlations.
In the phase-space region where Eq.~\eqref{eq:factorization}
holds, the \vn(\ptb) can be evaluated from the measured \vnn\ as:
\begin{eqnarray}
\vn(\ptb)=\frac{\vnn(\pta,\ptb)}{\vn(\pta)}=\frac{\vnn(\pta,\ptb)}{\sqrt{\vnn(\pta,\pta)}},
\label{eq:factorization_2}
\end{eqnarray}
where the relation $\vnn(\pta,\pta)=\vn^2(\pta)$ is used in
the denominator.
For most of the \tpc\ results in this analysis the \vn(\ptb) are
evaluated using Eq.~\eqref{eq:factorization_2}
with $0.5~\GeV<\pta<5~\GeV$.
The upper limit on \pta\ is chosen to exclude high-\pt\ particles,
which originate predominantly from jets.
 
\subsection{Template fits\label{sec:template}}
One drawback of the \tpc\ method is that in peripheral events with
low multiplicity or for \tpcs\ involving particles at high \pt, the
measured \vn\ can be biased by correlations arising from back-to-back
dijets that are not rejected by the $|\deta|>2$ requirement.
This issue is much more severe in smaller collision systems
such as \pPb\ and especially in \pp\ collisions, in which the
\tpc\ even at large $|\deta|$ is completely dominated by the back-to-back
dijet correlations.
To address this issue, a template-fitting procedure was developed
to measure the \vn\ in \pp\ and \pPb\
collisions~\cite{HION-2015-09,HION-2016-01,HION-2017-11}.
The template-fit method assumes that:
\begin{enumerate}
\item The shape of the dijet contribution to the \tpc\ does not change
from low- to high-multiplicity events, only its relative contribution
to the \tpc\ changes.
\item The \tpc\ for low-multiplicity events is dominated
by the dijet contribution.
\end{enumerate}
 
With the above assumptions, the correlation $\ctwophi$ in
higher multiplicity events is then described by a template fit,
\ctempl, consisting of two components: a scale factor, $F$,
times the correlation measured in low-multiplicity events,
\cperi, which accounts for the dijet-correlation, and a
genuine long-range harmonic modulation, \cridge:
\begin{eqnarray}
\ctempl &\equiv& F\cperi+\cridge\nonumber\\
&=     & F\cperi+G\left(1+2\Sigma_{n=2}^{\infty}v_{n,n}\cos(n\dphi)\right),\label{eq:template}
\end{eqnarray}
where the coefficient $F$ and the \vnn\ are fit parameters adjusted to
reproduce the $\ctwophi$.
The coefficient $G$ is not a free parameter but is fixed by the
requirement that the integrals of the $\ctempl$ and $C(\dphi)$
are equal.
There are two variations of the template-fit procedure, depending on how
the \cperi, used in Eq.~\eqref{eq:template}, is obtained.
In the first case, the entire correlation function in the
low-multiplicity interval is used as \cperi.
In the second case, only the modulated part of the correlation function
is used as \cperi\ and the unmodulated `pedestal' is removed.
The removal of the pedestal is done using the `zero yield at minimum' (ZYAM)
procedure~\cite{HION-2015-09}, which subtracts a pedestal from
the \cperi\ such that the value of \cperi\ is zero at its minimum.
The ZYAM-based method includes the additional assumption that only
the part of the low-multiplicity \tpc\ that is modulated in \dphi\ arises
from `non-flow' correlations.
In general, the standard and ZYAM-based template-fit measurements
in \pp\ and \pPb\ collisions give significantly different \vn\ values
at very low multiplicities.
But with increasing multiplicities, the difference between the \vn\ values obtained from
the two methods decreases~\cite{HION-2015-09,HION-2016-01}.
Since the \PbPb\ and \XeXe\ multiplicities are significantly larger than in \pp\ and \pPb\, it is expected that
the standard and the ZYAM-based template measurements yield
similar \vn\ values, except at very low multiplicities.
The difference between the \vn\ values obtained with/without the ZYAM procedure
then gives an estimate of the bias that the non-flow subtraction technique
induces in the measured \vn.

In this \papertype, the \cperi\ used in the template fits is constructed using
\pp\ events at $\sqs=5.02$~\TeV\ that have fewer than 20 reconstructed tracks passing the
selection requirements listed in Section~\ref{sec:datasets}.
Prior measurements using the template-fit method in \pp\ and \pPb\ collisions
used the low-multiplicity events from the same collision system to generate
the \cperi.
In the present analysis the choice of \pp\ reference is based on the following reasoning.
The template-fit method works better when the \cperi\ is dominated
by non-flow correlations.
Even in peripheral A+A collisions, there is still significant flow compared to $pp$ collisions,
thus using the \cperi\ constructed from \pp\ collisions at similar collision
energy, which has smaller flow-like correlations, is a better alternative.
Because of `jet quenching' effects present in heavy-ion
collisions~\cite{HION-2017-10,CMS-HIN-13-005,HION-2013-06,Abelev:2013kqa}, it is possible
that the assumption made in the template-fit procedure,
that the shape of the dijet correlation
in \dphi\ does not change from low- to high-multiplicity events, may not be valid.
The effects of jet quenching may bias the \vn\ measured using the template-fit method.
 
\subsection{Scalar product}
While the \tpc\ method relies on correlations between particle pairs
using only information from the ID to measure $v_n$, the SP~\cite{Adler:2002pu,Voloshin:2008dg}
measurement relies on correlations between the flow from energy deposits
measured in the FCal and from tracks in the ID.
Thus it allows measurements of the \vn\  with a larger gap in $\eta$
to strongly suppresses short-range correlations.
In fact, the larger $\eta$ gap not only suppresses short-range
correlations, it also suppresses correlations from back-to-back
dijets, as most of the dijets are at mid-rapidity.
 
The SP measurement is based on the construction of flow-vectors~\cite{Adler:2002pu,Voloshin:2008dg}
from tracks in the ID, and towers -- segmentations of the calorimeter of
approximate granularity $0.1\times0.1$ in $\eta$ and $\phi$ -- in the FCal.
The flow-vectors are defined as:~\footnote{The flow-vectors can be represented
as two-dimensional vectors, or equivalently as complex numbers, with the real and imaginary parts
of the complex number representing the $x$ and $y$ components of the flow-vector, respectively.
In this \papertype\ the complex-number notation is used.}
\begin{eqnarray*}
\kn=\frac{1}{\Sigma_jw_j}\sum_{j}w_j e^{in\phi_{j}},
\end{eqnarray*}
where $n$ is the harmonic order.
For the construction of $\kn$ from ID tracks, which is
labeled as $\qn$ in the following text,
$\phi_j$ is the azimuthal angle of the track,
and the weight $w_j$, which corrects for tracking performance,
is equal to $(1-f(p_{\mathrm{T}}^j,\eta^j))/\epsilon(p_{\mathrm{T}}^j,\eta^j)$.
The sum runs over a set of particles in a single event,
usually restricted to a region of the $\eta$--$\pT$ space.
For the estimate of $\kn$ from the FCal,
denoted by $\Qn$, the sum runs over the calorimeter
towers, with $\phi_j$ being the azimuthal position of the tower,
and the weight $w_j$ being the measured \et\ in the tower.

Using Eq.~\eqref{eq:single}, it can be shown that the flow-vectors $\qn$
are given by $v_n(\pt,\eta)e^{in\Phi_n}$.
However, due to statistical fluctuations arising from the finite number of tracks used in
measuring the $\qn$ event-by-event~\cite{HION-2012-10}, the measured $\qn$ fluctuate around
the true $v_n(\pt,\eta)e^{in\Phi_n}$, and can be written as: 
\begin{eqnarray}
\qn=v_n e^{in\Phi_n} + \qn^{\mathrm{fluc}},
\label{eq:flow_vec_2}
\end{eqnarray}
where $\qn^{\mathrm{fluc}}$ is a complex number representing event-by-event
statistical fluctuations.
Similarly the $\Qn$ can be written as:
\begin{eqnarray}
\Qn=V_n e^{in\Phi_n} + \Qn^{\mathrm{fluc}},
\label{eq:flow_vec_3}
\end{eqnarray}
where $V_n$ is used to denote the integrated \vn\ of particles in the calorimeter
acceptance together with the \ET\ response of the calorimeter folded in,
and $\Qn^{\mathrm{fluc}}$ represents statistical noise.

Due to the random orientation of the collision geometry from event to event,
the flow-vector averaged over many events should be equal to zero.
Additionally, the distributions for the real and imaginary parts of the
flow-vector should have identical widths.
However, due to nonuniform detector response in $\phi$, these conditions
may not be satisfied.
To correct for the nonuniform detector response, a procedure described in Ref.~\cite{HION-2016-06}
is applied, which ensures that the distributions of the real and imaginary parts of the
flow-vectors are centered at zero and have the same widths.

In the estimation of \vn\ from the SP method, four flow-vectors are
involved: $\Qn^P$ measured in the FCal at positive $\eta$, $\Qn^N$
measured in the FCal at negative $\eta$, and the corresponding
flow-vectors for charged particles measured in the ID, denoted by
$\qn^P$ for positive $\eta$ and $\qn^N$ for negative $\eta$.
The $\vn$ for $\eta<0$ from the SP method is then defined as
\begin{eqnarray}
\vn^{\mathrm{SP}}(\eta<0)=
\operatorname{Re} \frac{\langle \qn^{N}\Qn^{P*} \rangle}{\sqrt{\langle \Qn^{N}\Qn^{P*}\rangle}},
\label{eq:vnSP}
\end{eqnarray}
while for $\eta> 0$ the numerator is replaced by $\langle \qn^{P}\Qn^{N*}\rangle$.
The `*' denotes complex conjugate and the angular brackets
indicate an average over all events.
Equation~\eqref{eq:vnSP} can be understood by substituting expressions from Eq.~\eqref{eq:flow_vec_2}
and Eq.~\eqref{eq:flow_vec_3} for the flow-vectors, and noting that all terms
involving $\qn^{\mathrm{fluc}}$ and $\Qn^{\mathrm{fluc}}$ drop out when averaged over many events.
This gives:
\begin{eqnarray}
\vn^{\mathrm{SP}}(\eta<0)= \frac{\langle v^{N}_{n}V_{n}^{P} \rangle}{\sqrt{\langle V_{n}^{N}V_{n}^{P}\rangle}},
\label{eq:vnSP2}
\end{eqnarray}
where the superscripts $N$ and $P$ indicate whether
the quantities involved correspond to $\eta<0$ or $\eta>0$, respectively.
Equation~\eqref{eq:vnSP2} is similar to Eq.~\eqref{eq:factorization_2} for the \tpc\ measurements.
The final \vn\ from the SP method is obtained by averaging the results obtained for $\eta>0$
and $\eta<0$ in Eq.~\eqref{eq:vnSP}.
While Eq.~\eqref{eq:vnSP} explicitly uses the real part of the flow-vector product
to obtain the \vn, the imaginary part of the flow-vector product should be statistically
consistent with zero (cf.\ Eq.~\eqref{eq:vnSP2}).
Any statistically significant  nonzero value for the imaginary part of the flow-vector
product in Eq.~\eqref{eq:vnSP} is indicative of detector response effects that are not
corrected for in the measurement, and is typically used as a systematic uncertainty
of the measured \vn.

The \tpc\ and SP methods are very closely related;
both methods nominally measure the $\langle\sqrt{v_n^2}\rangle$,
where the average is taken over all particles and
events in the chosen \pt, $\eta$ and centrality range.
The \tpc\ method uses correlation between pairs of tracks
while the SP method utilizes correlations between tracks
and energy deposition in the calorimeters.
One advantage of the SP method in ATLAS
is that the larger rapidity gap between the FCal and
the ID, which is greater than 3.2 units, leads to larger
suppression of non-flow correlations as compared to the
\tpc\ method, where the minimum gap between the
reference and associated tracks is 2 units.
However, the SP method (in ATLAS) has some disadvantages:
for example, the larger $\eta$ gap biases the \vn\ measurements
if longitudinal flow decorrelations are present~\cite{CMS-HIN-14-008,CMS-HIN-15-008,HION-2016-04}.
The flow decorrelations increase with decreasing system-size~\cite{CMS-HIN-14-008,CMS-HIN-15-008,HION-2016-04},
and would affect flow measurements for \XeXe\ more than for \PbPb.
Additionally, for very low multiplicities, it is
very difficult to obtain reliable flow-vectors using
the calorimeters.
This is the reason why for small systems (\pp\ and \pPb) the flow
measurements are typically performed using the \tpc\ or
template-fit methods~\cite{HION-2015-09,HION-2016-01,CMS-HIN-16-010,HION-2017-11,Abelev:2012ola}.
 
\section{Systematic uncertainties\label{sec:systematics}}
The systematic uncertainties of the measured \vn\ are evaluated by
varying several aspects of the analysis.
Most of the uncertainties are common to the SP and \tpc\
methods and are discussed together.
Since the template-fit measurements start with the \tpcs,
all systematic uncertainties related to the \tpc\ method
also affect the measurements based on the template fit.
The template-fit measurements have an additional uncertainty
related to the choice of the peripheral reference,
which is discussed below.
The SP method is used to measure harmonics $v_2$--$v_6$,
while the \tpc\ and template-fit methods are used to
measure harmonics $v_2$--$v_5$.
This difference is due to the fact that the \tpc\ and template-fit measurements have
large systematic uncertainties attributed to the pair acceptance and
peripheral reference choice for $v_6$, which make a significant measurement
not feasible.
The uncertainties for two representative centrality and \pt\ ranges
are summarized in Table~\ref{tab:uncSP}.
The following sources of uncertainties are considered:
 
\begin{itemize}
\item \textbf{Track selection:}
The tracking selection cuts control the relative contribution of
genuine charged particles and fake tracks entering the analysis.
The stability of the results with respect to the track selection is evaluated by
varying the requirements imposed on the reconstructed tracks,
and including the resulting variation in the \vn\ as a
systematic uncertainty.
The two sets of variations termed loose and tight
in Section~\ref{sec:datasets} are used for this purpose.
At low \pT (0.5--0.8~\GeV) the variation in the \vn\ obtained from this
procedure is most significant in the most central events, typically 5\%, as the
fake-track rate is largest in this region of phase space.
For higher \pT and less central events, changing the set of tracks
used in the analysis has less influence on the measurement.

\item \textbf{Tracking efficiency:}
As mentioned in Section~\ref{sec:methodology}, the tracks are weighted by $(1-f)/\epsilon$
when calculating the \vn\ to account for the impact of the
tracking efficiency.
Uncertainties in the efficiency, resulting from e.g. uncertainty of
the amount of material in the detector, are propagated into
the measured \vn.
This uncertainty is evaluated by varying the efficiency up and down
within its uncertainties ($\sim\pm$3\%) in a \pT-dependent manner and re-evaluating
the \vn.
This contribution to the overall uncertainty is very small and
amounts to less than 1\% on average for the \pt-integrated \vn,
and is negligible for the \pt-differential \vn.
This is because the change of efficiency largely cancels out in the
differential $\vn(\pT)$ measurement, and for \vn\ integrated over
\pT, the low-\pT particles dominate the measurement.
The uncertainty does not change significantly with centrality or with
the harmonic order.

\item \textbf{Centrality determination:}
The centrality definitions used to classify the events into
centrality percentiles have a $\sim$1\% uncertainty associated
with them.
This uncertainty arises from uncertainties in the fraction of the
inelastic \XeXe\ cross-section accepted by the triggers used in this analysis.
The impact of the uncertainty from the centrality definition on the \vn\
is evaluated by varying the centrality interval definitions by 1\%,
re-evaluating the \vn\ and including the variation in the \vn\ as
a systematic uncertainty.
The impact on all harmonics over the 0--50\% centrality range is found to be within 1\%.
For more-peripheral events, this number varies between 1\% and 5\% depending on the
\pt, centrality and harmonic order $n$.
 
\item \textbf{MC closure:}
The MC closure test consists of comparing the $\vn^\mathrm{true}$
obtained directly from the MC generated particles,
and the $\vn^\mathrm{reco}$ obtained by applying
the same analysis procedures to the MC sample as to the data.
The analysis of MC events is done to evaluate the contributions of
effects not corrected for in the data analysis.
Systematic differences seen between the \vn\ of the generated particles and
reconstructed particles are used to correct the \vn\ measured in the
data and, conservatively, also included as a systematic uncertainty.
This uncertainty is at the level of a few percent over the
0.5--0.8~\GeV\ \pt\ range and the 0--20\% centrality range, and reaches
5\% for $\pt\sim0.5$~\GeV\ in the 0--5\% centrality interval.
It is negligible elsewhere.
 
\item \textbf{Event-mixing:}
As explained in Section\,\ref{sec:2pc}, the \tpc\ analysis uses the
event-mixing technique to estimate and correct for the detector
acceptance effects.
Potential systematic uncertainties in the \vn\ due to the residual
pair-acceptance effects, which are not corrected for by the mixed
events, are evaluated by varying the multiplicity and $z$-vertex
matching criteria used to make the mixed-event distributions,
following procedures from Ref.\,\cite{HION-2011-01}.
The resulting uncertainty in the \vtwo--\vfive\ is 1--3\%
for most of the centrality and \pt\ ranges considered
in this \papertype.
However, at high-\pt\ and for the harmonic orders $n\geq4$,
this uncertainty can become at large as 10\%.
This uncertainty only contributes to the \vn\ values measured by the
\tpc\ and template-fit methods.
 
\item \textbf{Choice of peripheral reference:}
The template-fit procedure uses \pp\ events at $\sqs=5.02$~\TeV\
with less than 20 reconstructed tracks to build $\cperi$.
The choice of 20 tracks is partially motivated by the fact that the mean
multiplicity of minimum-bias \pp\ events is close to 20 tracks.
To test the stability of the \vn\ with respect to our choice of the peripheral
reference, the analysis is repeated with an alternative $\cperi$
constructed from \pp\ events with 0--20, 10--20 and 10--30
reconstructed tracks and the change in the template-\vn\ values
is included as a systematic uncertainty.
This uncertainty is within $\sim$4\% over the 0--50\% centrality range
and for $\pt<$~4~\GeV, but increases considerably and can become as
large as 30\% for more-peripheral events or at higher \pt.
This uncertainty only contributes to the \vn\ values measured by the
template-fit method.

\item \textbf{$\eta$ asymmetry:}
Due to the symmetry of the Xe+Xe collision system the event-averaged
$\langle\vn(\eta)\rangle$ and $\langle\vn(-\eta)\rangle$
are expected to be equal.
Any difference between the event-averaged \vn\ at $\pm\eta$ arises
from residual detector nonuniformity.
The difference between the \vn\ values measured in opposite
hemispheres is treated as a systematic uncertainty quantifying
imperfect detector performance.
This uncertainty only contributes to the \vn\ values measured with the
SP method.
The value of this uncertainty depends on the harmonic order.
It is less than $1\%$ for $v_2$ and increases to $\sim$5\% for $v_6$.
For the \tpc\ method, the residual nonuniformity is estimated by
variations in the event-mixing procedure.

\item \textbf{Residual sine term:}
The imaginary part of the flow-vector product in Eq.~\eqref{eq:vnSP}
should be consistent with zero.
Any systematic deviation of the imaginary part from zero
indicates the presence of residual detector response effects.
The ratio of the imaginary part of the flow-vector product to its
real part (Eq.~\eqref{eq:vnSP}) is therefore included as a contribution to the
systematic uncertainty.
The contribution from this source is  $\leq2\%$ in most of
the phase space, while for the higher harmonics ($n=5$)
it can reach 10\%.
This uncertainty is only relevant for the \vn\ values measured by the
SP method.
 
\end{itemize}
 
\begin{table}
\caption{
The relative contributions (in percent) to the systematic uncertainty of
\vn\ in two selected bins of centrality.
The contributions are expressed as a percentage of the measured \vn\
and are rounded up to two significant digits.
Items 1--4 are common to all methods.
Item 5 is specific to the \tpc\ and template-fit methods,
item 6 is specific to the template-fit method,
and items 7--8 are specific to the SP method only.
The \tpc\ and template-fit methods are used to measure harmonics \vtwo--\vfive,
while the SP method is used to measure harmonics \vtwo--\vsix.
\label{tab:uncSP}
}
\begin{center}
\begin{tabular}{|l c|c c|c c|c c|}
\hline
\multirow{2}{1.3cm}{\centering Uncertainty sources } &
\multirow{2}{1.5cm}{\centering Harmonic order} &
\multicolumn{2}{c}{ 5--10\% centrality} & \multicolumn{2}{|c|}{40--50\% centrality}   \\
& & 0.8--1~\GeV\ & 6--8~\GeV\     &  0.8--1~\GeV\ & 6--8~\GeV\   \\
\hline
1. Track selection          & \vtwo--\vsix       & 1.5  & 1  & 0.5 & 1   \\
2. Tracking efficiency      & \vtwo--\vsix       & 1    & 1  & 1   & 1   \\
3. Centrality determination & \vtwo--\vsix       & 1    & 1  & 1   & 1   \\
4. MC closure               & \vtwo--\vsix       &<0.1  &<0.1 &<0.1&<0.1 \\
\hline
\multirow{4}{*}{ \centering 5. Event-mixing}
& \vtwo                 &  1& 1        &    1&  1      \\
& \vthree               &  1& 1        &    1&  3.5    \\
& \vfour                &  1& 6        &    1&  6      \\
& \vfive                &  4   &10        & 4   & 10      \\
\hline
\multirow{4}{*}{ \centering 6. Peripheral reference}
& \vtwo                 &  <0.5& 1.5      &     0.5&  1.5    \\
& \vthree               &   0.7& 3.5      &     0.9&  6      \\
& \vfour                &   1.0& 14       &     2.0&  18      \\
& \vfive                &   5  & 30       &     5  &  30      \\
\hline
\multirow{4}{*}{ \centering 7. $\eta$ asymmetry}
& \vtwo         & 1 & 1 & 1 & 1\\
& \vthree       & 2 & 2 & 2 & 2\\
& \vfour        & 3 & 3 & 3 & 3\\
& \vfive--\vsix & 5 & 5 & 5 & 5\\
 
\hline
\multirow{3}{*}{ \centering 8. Residual sine term}
& \vtwo-\vthree   & 2 & 2 & 2 & 2\\
& \vfour          & 4 & 4 & 4 & 4\\
& \vfive--\vsix   & 10 & 10 & 10 & 10\\
\hline
\end{tabular}
\end{center}
\end{table}

\section{Results\label{sec:results}}
\subsection{Two-particle correlations and template fits}
\begin{figure}[H]
\centering
\includegraphics[width=1.00\linewidth]{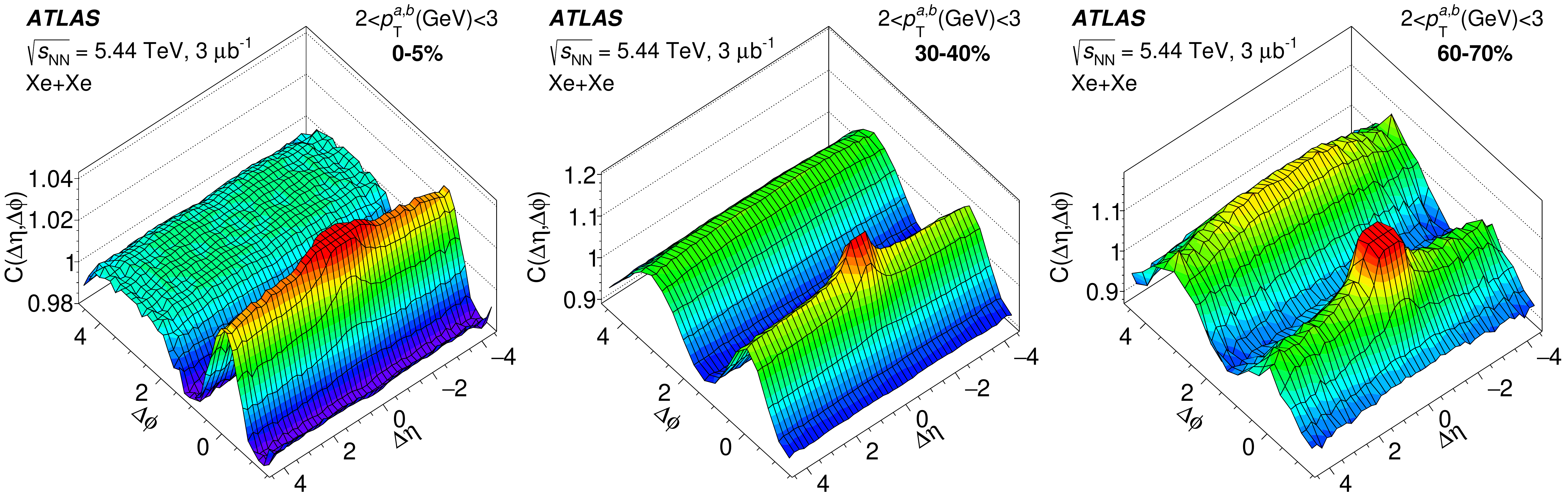}
\caption{
Two-particle correlations in $\Delta\eta$--$\Delta\phi$ for $2~\GeV<p_{\mathrm {T}}^{\mathrm{a,b}}<3~\GeV$.
The correlations are shown for the 0--5\% (left), 30--40\% (center) and 60--70\% (right) centrality intervals.
The distributions are truncated to suppress the peak at $\deta=\dphi=0$
to show the long-range correlation in greater detail.
The $\Delta\eta$ axis is also truncated at $\Delta\eta=\pm4.6$ to avoid
large statistical fluctuations present at the edge of the $\Delta\eta$ acceptance.
}
\label{fig:2d_2pcs}
\end{figure}

Figure~\ref{fig:2d_2pcs} shows the two-dimensional 2PCs
in $\Delta\eta$--$\Delta\phi$ for several centrality intervals
for $2~\GeV<\ptab<3~\GeV$, where $\pta$ and $\ptb$ label the $\pt$ ranges of the
reference and associated particles used in the correlation.
In all cases a peak is seen in the correlation at $(\deta,\dphi)\sim(0,0)$.
This peak arises from short-range correlations such as
jet-fragmentation, resonance decays or Hanbury Brown and Twiss (HBT) correlations~\cite{Lisa:2005dd}.
The long-range (large $\deta$) correlations are the result of the global
anisotropy of the event and are the focus of the study in this analysis.
Thus, all further \tpc\ results are discussed with the $\deta>2$ requirement.
 
\begin{figure}[H]
\centering
\includegraphics[width=1.00\linewidth]{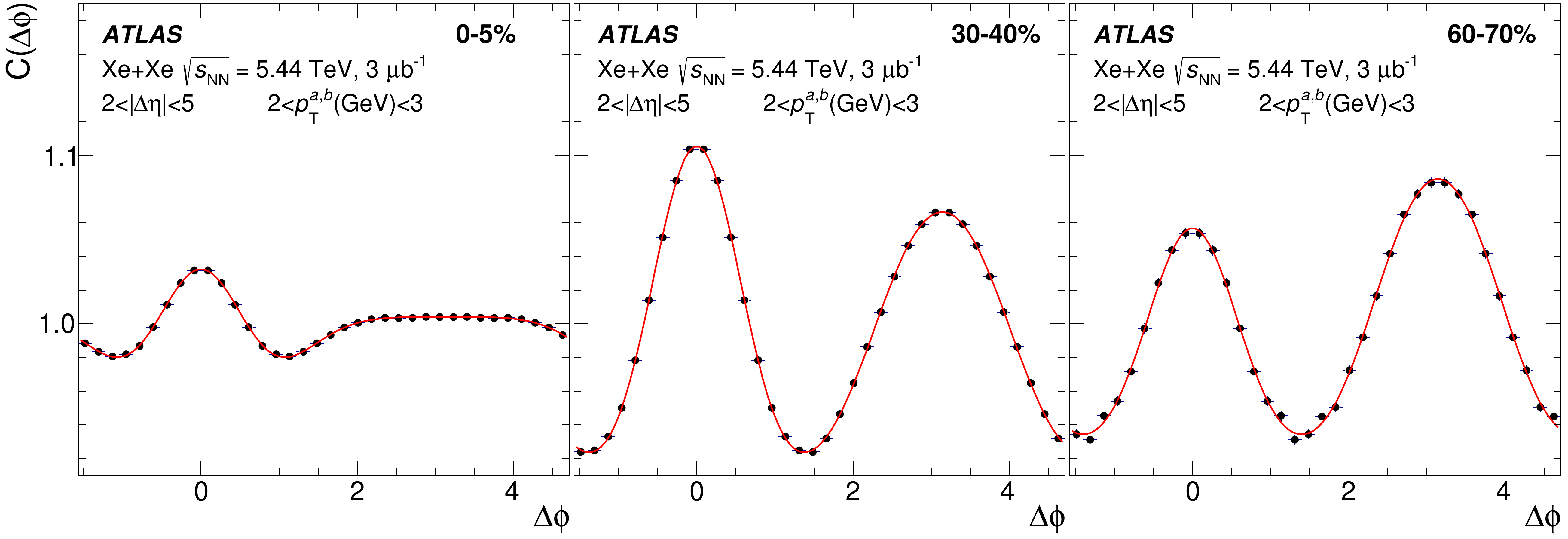}
\caption{
Two-particle correlations in $\Delta\phi$ for $2<|\Delta\eta|<5$ and $2~\GeV<p_{\mathrm {T}}^{\mathrm{a,b}}<3~\GeV$.
The correlations are shown for the 0--5\% (left), 30--40\% (center) and 60--70\% (right) centrality intervals.
The lines represent Fourier fits to the correlation functions (Eq.~\eqref{eq:2pc_fourier})
that include harmonics up to $n=5$.
}
\label{fig:1d_2pcs_1}
\end{figure}

Figure~\ref{fig:1d_2pcs_1} shows corresponding one-dimentional 2PCs along $\dphi$
for $2~\GeV<\ptab<3~\GeV$ and for several centrality intervals.
The line is a Fourier fit to the correlation
(Eq.~\eqref{eq:2pc_fourier}) that includes harmonics up to $n=5$.
The $y$-axis ranges for the different panels are kept identical so that the
modulation in the correlation across the different centralities can be
easily compared.
It is seen that the modulation in the correlation is smallest in the most
central events (left panel), increases towards mid-central events
and then decreases slightly.
This roughly follows the centrality dependence of most $v_{n,n}$,
especially the $v_{2,2}$.
 
\begin{figure}[H]
\centering
\includegraphics[width=1.00\linewidth]{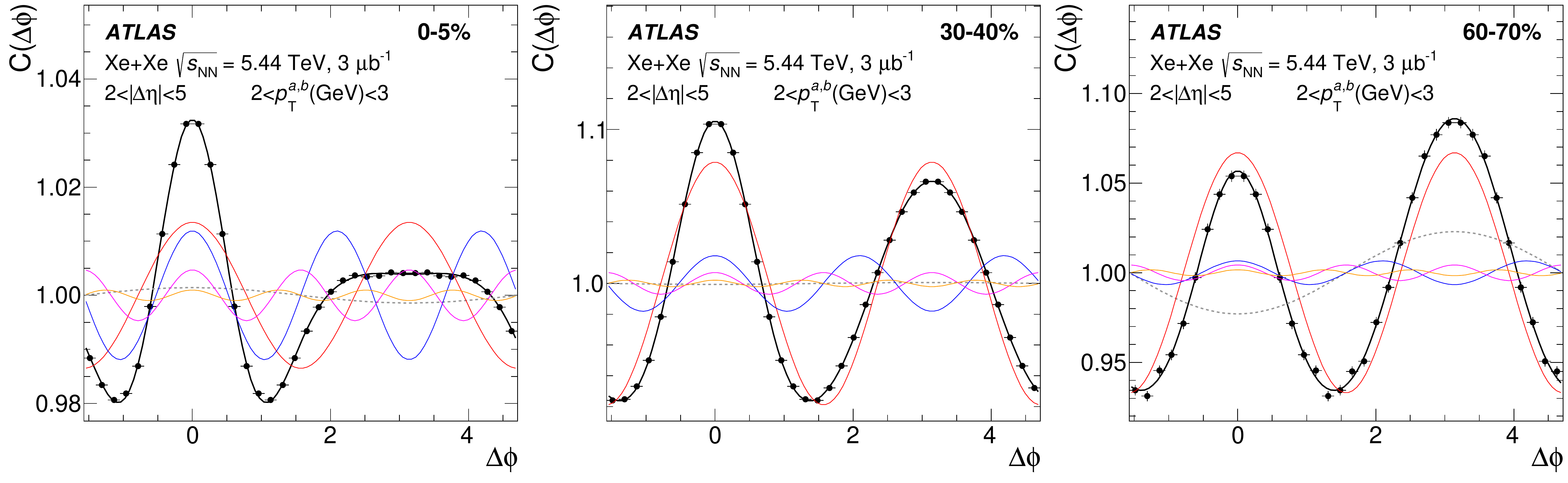}
\caption{
Two-particle correlations in $\Delta\phi$ for $2<|\Delta\eta|<5$ and $2~\GeV<p_{\mathrm {T}}^{\mathrm{a,b}}<3~\GeV$.
The correlations are shown for the 0--5\% (left), 30--40\% (center) and 60--70\% (right) centrality intervals.
Also shown is a Fourier fit to the correlation (Eq.~\eqref{eq:2pc_fourier}) that
includes harmonics $n$ up to 5 (black line).
The colored lines show the contribution of the $v_{2,2}$ (red), $v_{3,3}$ (blue), $v_{4,4}$ (magenta)
and $v_{5,5}$ (orange), and can also be identified by the number of peaks in $\Delta\phi$.
The dotted line indicates the $v_{1,1}$.
Each panel represents a different centrality interval.
}
\label{fig:1d_2pcs_1a}
\end{figure}
 
Figure~\ref{fig:1d_2pcs_1a} shows the same correlation functions but with a reduced
$y$-axis range to make it easier to observe the features
of the correlation.
The Fourier fit is indicated by the thick black line and the contributions of the
individual $v_{n,n}$ are also shown.
In the most central 0--5\% collisions (left panel of Figure~\ref{fig:1d_2pcs_1a}) the $v_{2,2}$--$v_{4,4}$ are of
comparable magnitude, but in other centralities, where the
average collision geometry is elliptic, the $v_{2,2}$ is
significantly larger than the other $v_{n,n}$ ($n\geq3$).
In central events, the away-side peak (at $\dphi=\pi$) is also much broader due to the
comparable magnitudes of the higher-order harmonics and $v_{2,2}$, while in mid-central
events (30--40\% centrality in Figure~\ref{fig:1d_2pcs_1a}) the near ($\dphi=0$) and away-side peaks
are more symmetric as the $v_{2,2}$ dominates.
In central and mid-central events, the near-side peak is larger than the
away-side peak.
However, beyond 60\% centrality (right panel of Figure~\ref{fig:1d_2pcs_1a}) the away-side peak becomes larger, indicating
the presence of a large negative $v_{1,1}$ component.
This negative $v_{1,1}$ component in the peripheral 2PCs largely arises
from dijets and its contribution to the correlation function increases
in peripheral events~\cite{HION-2011-01,HION-2016-01}.
While the near-side jet contribution is rejected by the $|\deta|>2$ requirement,
the away-side jet is not localized in $|\deta|$ and cannot be rejected entirely.
The presence of the away-side jet produces a large negative $v_{1,1}$ and also
affects the other harmonics.
Over this centrality range, the $v_{n,n}$ are strongly biased by dijets,
especially at higher $\pt$.
The presence of the jets also results in the breakdown of the factorization
relation (Eq.~\eqref{eq:factorization_2}).
This is discussed in more detail in Section~\ref{subsec:factorization}.

\begin{figure}[H]
\centering
\includegraphics[width=1.00\linewidth]{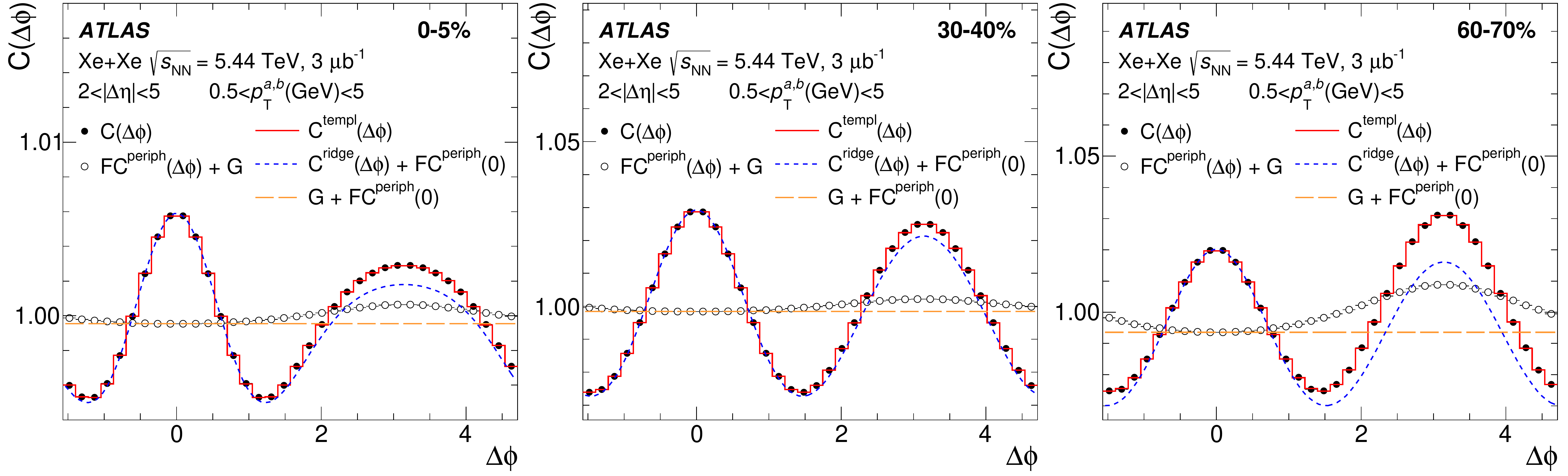}
\caption{
Template fits to the 2PCs in Xe+Xe collisions.
The fits are shown for the 0--5\% (left), 30--40\% (center) and 60--70\% (right) centrality intervals.
The plots are for $0.5~\GeV<p_{\mathrm {T}}^{\mathrm{a,b}}<5~\GeV$.
The solid points indicate the measured 2PC, the open points and
curves show different components of the template (see legend)
that are shifted along the $y$-axis by $G$ or by
$FC^{\mathrm{periph}}(0)$, where necessary, for presentation.
The plots correspond to the template-fit method without
using ZYAM subtraction on the $pp$ reference.
The template-fit includes harmonics $n$ up to 5.
}
\label{fig:Fit_pta7_ptb49_ch2_deta01_peri00_ppReference}
\end{figure}

Figure~\ref{fig:Fit_pta7_ptb49_ch2_deta01_peri00_ppReference}
shows examples of template fits
to the measured \tpcs\ (Eq.~\eqref{eq:template}),
that includes harmonics $n$ up to 5.
The figure shows the template fits without a ZYAM subtraction
performed on the \pp\ reference.
However, in the plots presented in
Figure~\ref{fig:Fit_pta7_ptb49_ch2_deta01_peri00_ppReference},
the template fits look identical with and without ZYAM as explained in the following.
The ZYAM method sets the minimum of the \cperi\ term
to zero by removing a pedestal equal to
$C^{\mathrm{periph}}(0)$, since the minimum of \cperi\
occurs at $\dphi=0$.
The template fits then transfer this pedestal to the $G$
term of the \cridge.
However, the components of the template fit are
shifted vertically by the pedestal of the other term,
i.e., the \cridge\ is plotted as `$\cridge +FC^{\mathrm{periph}}(0)$'
and the \cperi\ is plotted as `$F\cperi+G$',
where $G$ and $FC^{\mathrm{periph}}(0)$ are the pedestals of the
two components.
This shift in the plotting undoes the shift in the pedestals of the
two components.
In Figure~\ref{fig:Fit_pta7_ptb49_ch2_deta01_peri00_ppReference}
the fits are shown for a central (0--5\%), a mid-central (30--40\%)
and a peripheral (60--70\%) centrality interval.
The relative difference between \ctempl\ and \cridge\ is largest for
the peripheral centrality interval and decreases for mid-central events.
This is indicative of larger contributions from non-flow correlations
in peripheral collisions than in mid-central collisions.

\subsection{Factorization of \vn\ for \tpc\ and template-fits \label{subsec:factorization}}
 
\begin{figure}[H]
\centering
\includegraphics[width=0.9\linewidth]{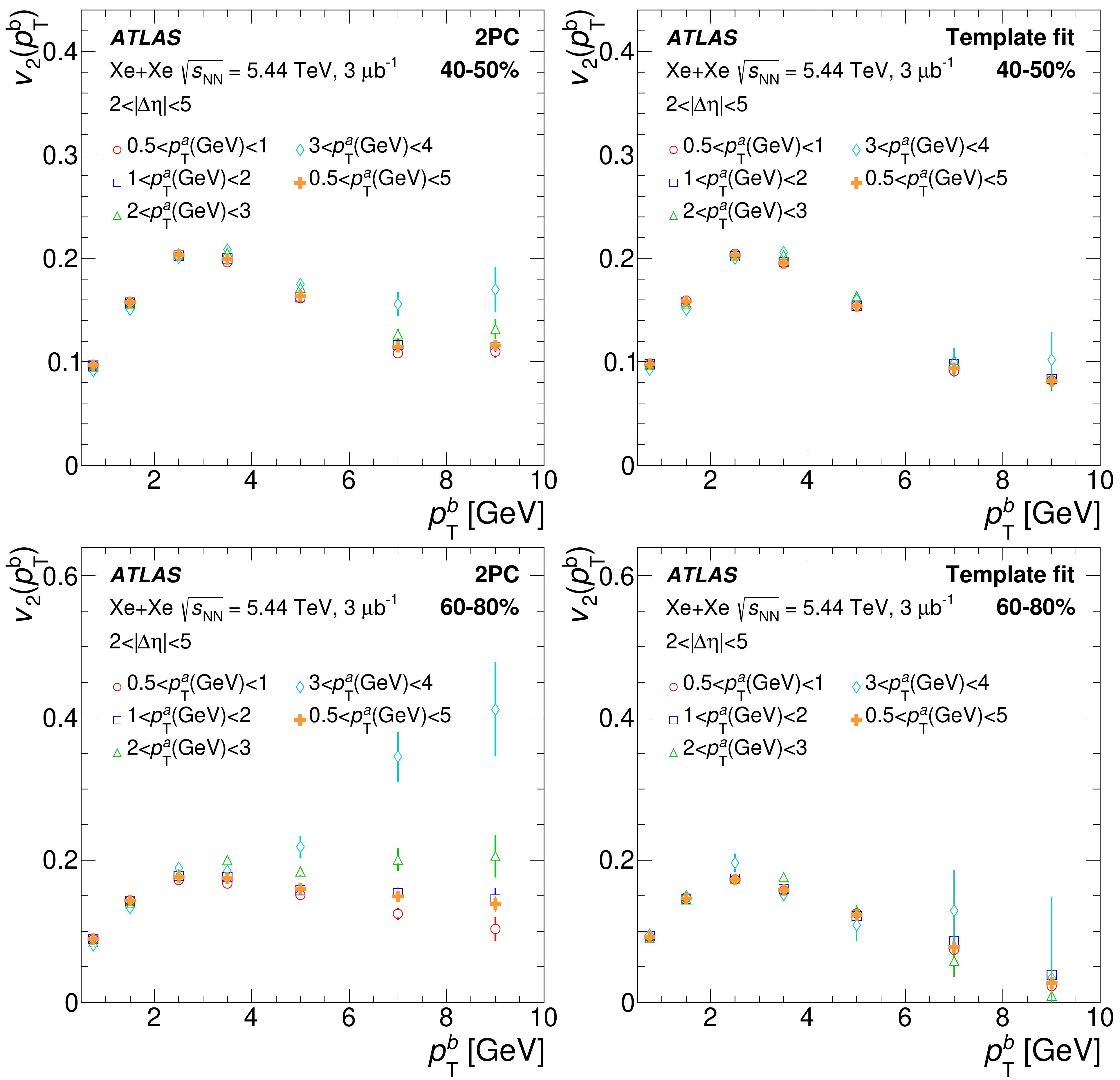}
\caption{
The demonstration of factorization as a function of $p_{\mathrm {T}}$ for $v_2$ in the
40--50\% centrality interval (top row) and 60--80\% centrality interval (bottom row).
The left panels show the results for the 2PC method and the right panels
show the results for the template-fit method without ZYAM.
The error bars indicate statistical uncertainties only.
}
\label{fig:factorization_vn_ptdep}
\end{figure}

Figure \ref{fig:factorization_vn_ptdep} shows how well the factorization
of the \vn\ (Eq.~\eqref{eq:factorization_2}) works for the \tpc\ and
template-fit \vn\ measurements.
The plots show the $\vtwo(\ptb)$ obtained using Eq.~\eqref{eq:factorization_2}
in the 40--50\% and 60--80\% centrality ranges,
for several \pta\ intervals: 0.5--1~\GeV, 1--2~\GeV, 2--3~\GeV, 3--4~\GeV\ and 0.5--5~\GeV.
If factorization worked, then the obtained values of $\vtwo(\ptb)$ would be
independent of the choice of \pta.
For the \tpc\ measurements, factorization breaks down at large values of
\pta\ or \ptb.
The exact values of \pt at which the breakdown occurs depend
on the centrality.
The breakdown in factorization occurs mostly because of non-flow effects such as jets.
In mid-central events the $v_{2}$ is largest, thus the bias from non-flow effects
is relatively weak and the factorization holds until much higher
$\pt$ values than in other centrality ranges.
In peripheral events, with the decreased multiplicity, the bias from
non-flow effects becomes larger; this, coupled with the decreased $v_{2}$,
results in larger or earlier onset (in $\pt$) of factorization breakdown.
In general, the factorization holds better for $v_{3,3}$ and $v_{4,4}$
than for $v_{2,2}$.
This is because the factorization breakdown occurs mostly because of the
back-to-back dijets, not rejected by the $\deta$ cut, which have a larger
impact on the $v_{2,2}$ than the $n\geq3$ harmonics~\cite{HION-2011-01,HION-2016-01}.
These observations are similar to those made in previous $v_{n}$
measurements in Pb+Pb collisions at $\snn=5.02$~\TeV~\cite{HION-2016-06}.
However, quantitatively, the breakdown in the \XeXe\ case is somewhat
larger than in \PbPb\ collisions at the same centrality,
due to smaller multiplicities.
On the other hand, for the template-fit measurements the factorization works
much better for higher \pt\ and for more-peripheral events, indicating
a significant reduction of the bias from dijet correlations.
 
For all remaining plots involving the \tpc\ or template-fit measurements,
the default \pta\ interval used for the \vn\ measurement
is $0.5~\GeV<\pta<5$~\GeV.
Over this $\pt$ interval the factorization holds quite well for the \tpc\
method for most of the centrality ranges.

\subsection{\pt\ dependence of \vn}
 
\begin{figure}
\centering
\includegraphics[width=0.9\linewidth]{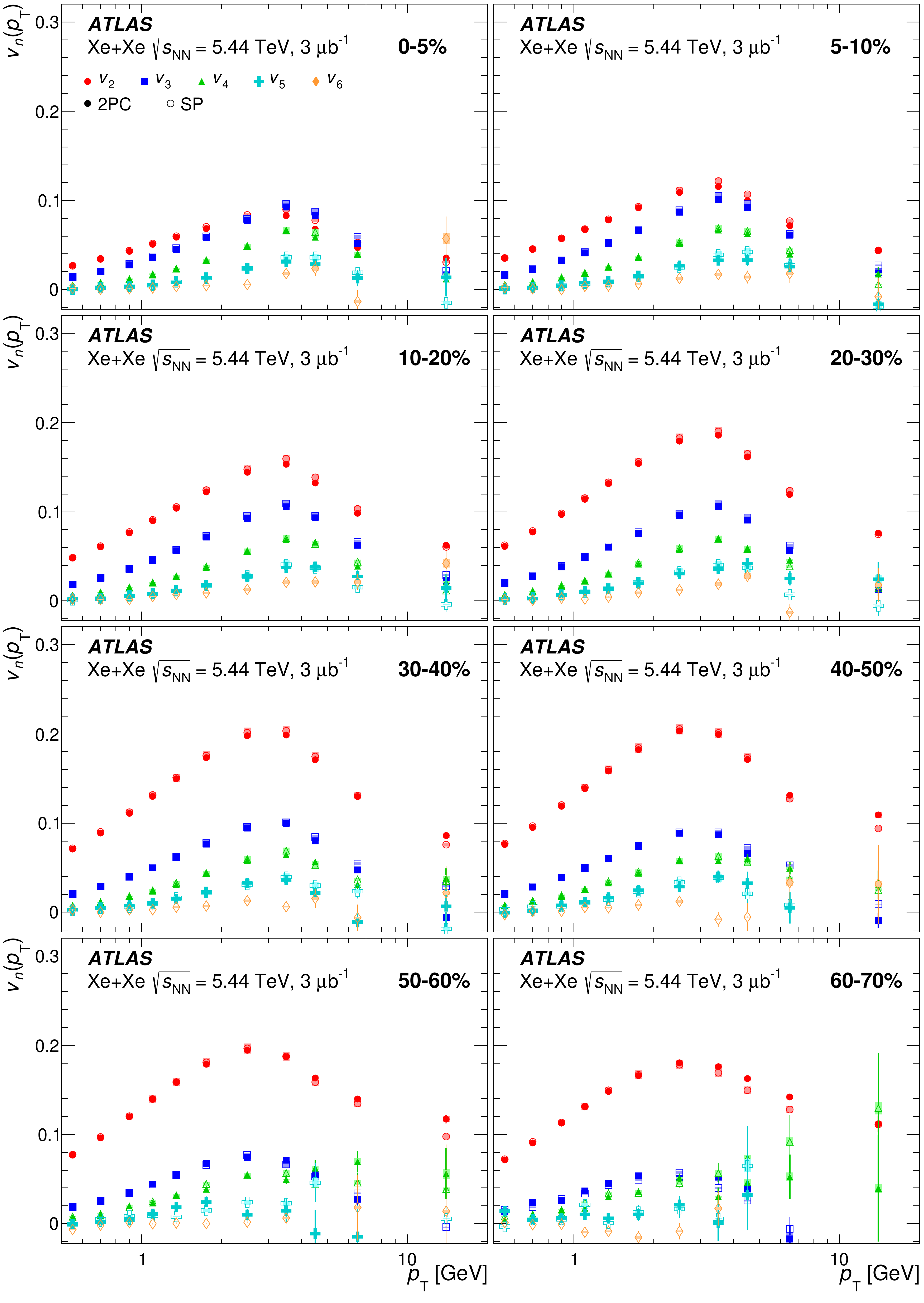}
\caption{
The \pt\ dependence of the $v_{n}$ obtained with the 2PC and SP methods.
Each panel represents a different centrality interval.
The solid markers show the results obtained using the 2PC method
and the open markers show the results obtained from the
SP method.
The vertical error bars and shaded bands indicate statistical
and systematic uncertainties, respectively, and are often
too small to be visible.
The \tpc\ results are shown for harmonics 2--5, and the
SP results are shown for harmonics 2--6.
}
\label{fig:XeXe_vn_ptDept_2PC_and_SP_full_aux}
\end{figure}

Figure\,\ref{fig:XeXe_vn_ptDept_2PC_and_SP_full_aux} shows the \pt\ dependence
of the \vn\ obtained using the \tpc\ and SP methods.
Each panel shows the measurements for a different centrality interval.
In general the \tpc\ and SP results are quite comparable, although small differences
can be seen in the most central events.
However, in peripheral events (50--60\% and 60--70\% centrality classes) and for $\pt>4$~\GeV\
the \vtwo\ from the \tpc\ method gives systematically higher values.
This increase arises due to bias from dijets, and is further investigated below.

\begin{figure}
\centering
\includegraphics[width=0.9\linewidth]{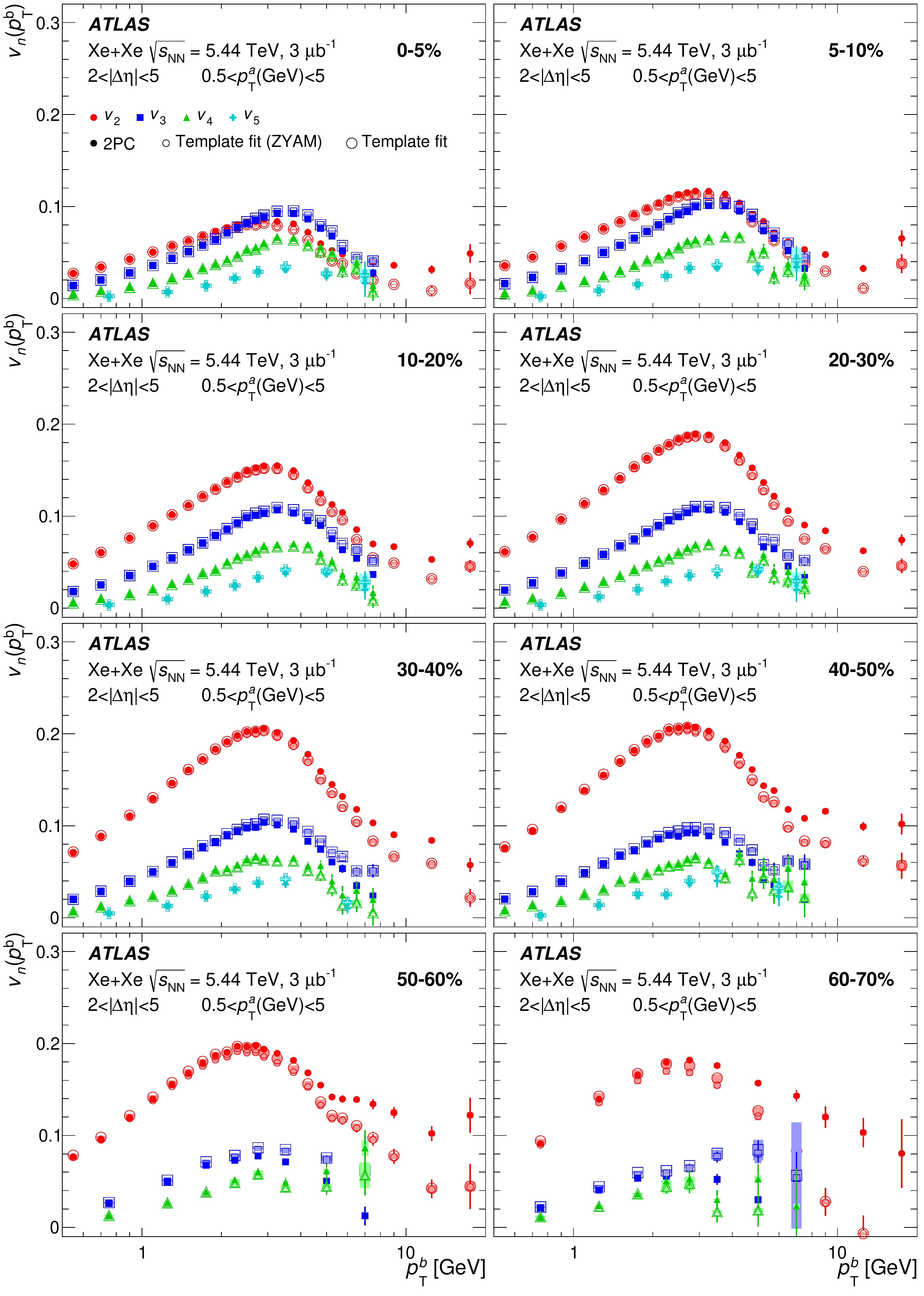}
\caption{
The $p_{\mathrm {T}}^{\mathrm{b}}$ dependence of the $v_{n}$ obtained with the 2PC and template-fit measurements.
Each panel represents a different centrality interval.
The solid markers show the results obtained using the 2PC analysis.
The small and large open markers show the results obtained from the
template fit with and without ZYAM, respectively.
The vertical error bars and shaded bands indicate statistical
and systematic uncertainties, respectively, and are often
too small to be visible.
}
\label{fig:XeXe_vn_ptDept_template_full}
\end{figure}

Figure\,\ref{fig:XeXe_vn_ptDept_template_full} compares the \pt\
dependence of the \vn\ obtained using the \tpc\ method, and the
template-fit measurement with/without the ZYAM subtraction.
The measurements are shown in finer \pt\ intervals, thereby allowing
examination of the \pt\ dependence and the non-flow bias in detail.
For the harmonic $v_{5}$ the results are truncated
for centralities more peripheral than the 40--50\% interval,
beyond which the statistical errors are typically too large to study
the \pt\ dependence.
For all the methods of measurement, the following trends are observed:
the \vn\ values increase at low \pt\ and reach a maximum between 2~\GeV\ and 4~\GeV, and
then decrease.
For nearly all centralities, the \vn\ follow the trend
$v_{2}\gg v_{3}>v_{4}>v_{5}$.
This hierarchy is violated in the most central 0--5\% collisions, where the
$v_{3}$ and $v_{4}$ values at high-enough \pt\ become larger than the $v_{2}$ values.
These trends are the same as those observed in prior \vn\ measurements
in \PbPb\ collisions~\cite{HION-2011-01,HION-2016-06}.
For the \tpc\ measurements, it is observed that at high \pt (the last two \pt\ intervals shown),
the \vtwo\ again increases with \pt.
This feature can be explained as a bias from dijets, which dominate the
\tpc\ at high \pt\ especially in peripheral events.
The back-to-back nature of the dijet correlations enhances (suppresses)
the measured values of the \vn\ for even-order (odd-order) harmonics.
This trend is to some extent also seen in the SP measurements
shown previously in Figure~\ref{fig:XeXe_vn_ptDept_2PC_and_SP_full_aux}.
This bias is considerably reduced in the measurements based on the template-fit method.
It is interesting to note that the template-fit measurements with or without ZYAM
give similar results for the \vn.
As mentioned earlier, the differences between the \vn\ values obtained
with/without the ZYAM procedure can give an estimate of possible biases
in the non-flow subtraction.
Thus, similar values obtained from the two template-fit methods, which make
different assumptions about the \cperi\ (Eq.~\eqref{eq:template}),
give more confidence in the robustness of the non-flow subtraction.

\subsection{Centrality dependence of \vn}
\begin{figure}
\centering
\includegraphics[width=0.9\linewidth]{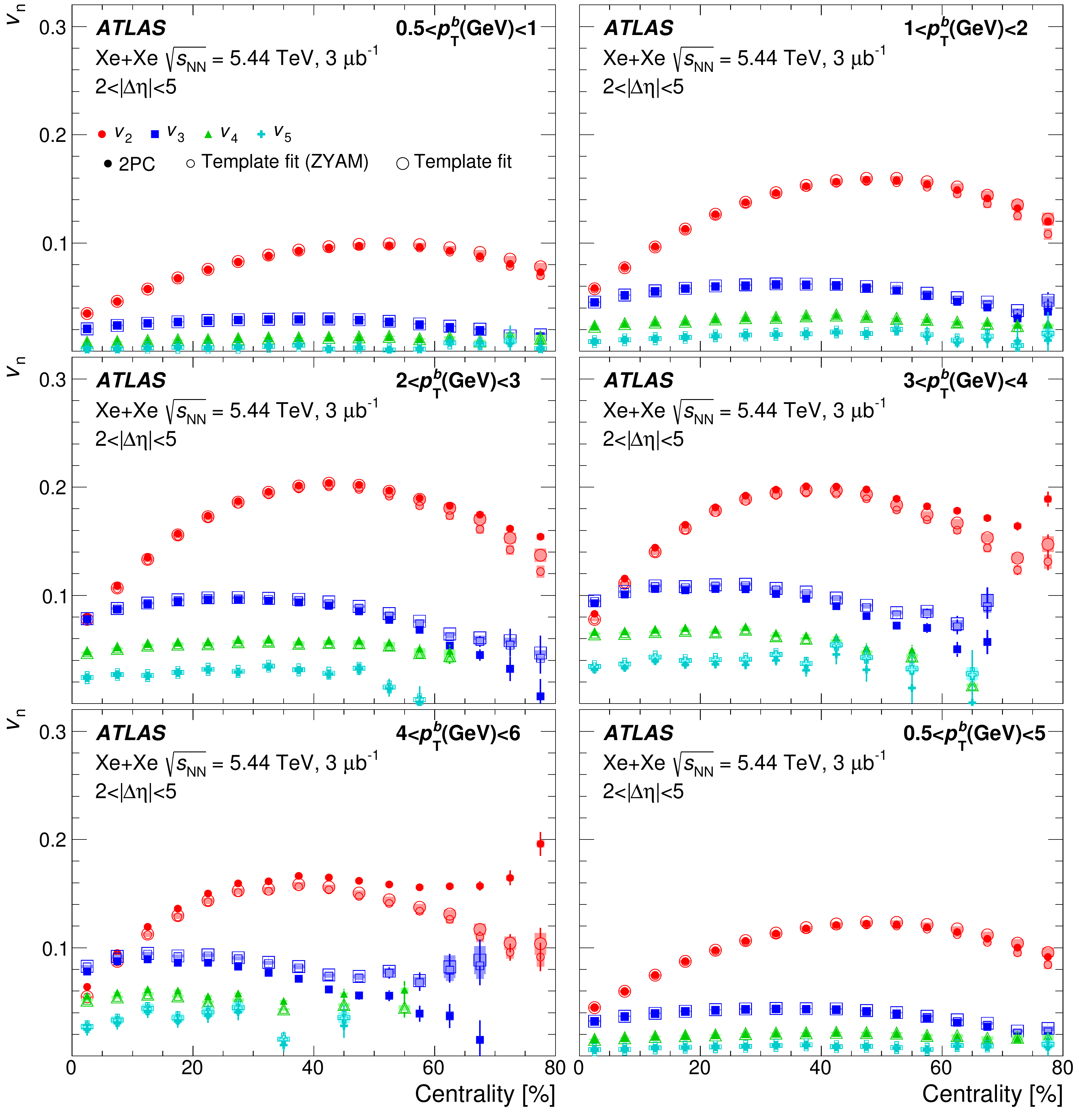}
\caption{
The centrality dependence of the $v_n$ obtained with the 2PC and template-fit methods.
The solid points represent the 2PC $v_n$.
The large open points represent the template fit $v_n$ without ZYAM.
The small open points represent the template fit $v_n$ with ZYAM.
Each panel shows the results for a different $p_{\mathrm {T}}$ interval.
The error bars and shaded bands indicate statistical and systematic uncertainties,
respectively.
}
\label{fig:vn_cent}
\end{figure}

Figure~\ref{fig:vn_cent} shows the centrality dependence of the \vn\
obtained from the \tpc\ and template-fit methods.
The different panels represent different \pt\ intervals.
Results from the template-fit method with/without a ZYAM
procedure on the reference and from the \tpc\ method are shown.
For the \tpc\ measurements and for $\pt<3~\GeV$ -- shown in the first three panels --
the \vn\ values increase from central to mid-central events,
reach a maximum between 40\% and 60\% centrality, and
then decrease.
The variation with centrality is largest for the
\vtwo\ owing to the large change in the second-order
eccentricity~\cite{Qiu:2011iv} from central to mid-central collisions.
These features are consistent with those observed in
other heavy-ion collisions~\cite{Back:2004je,Adcox:2004mh,Arsene:2004fa,HION-2011-05,HION-2011-01,HION-2016-06,Adams:2005dq,PhysRevLett.105.252302,CMS-HIN-10-002}.
The effects of the non-flow bias seen in Figure~\ref{fig:XeXe_vn_ptDept_template_full}
can also be observed more clearly in Figure~\ref{fig:vn_cent},
where for the 4--6~\GeV\ \pt\ interval the \tpc\ \vtwo\ values increase over
the 60--80\% centrality range.
The template-fit \vtwo\ on the other hand does not show this increase.
In general, for centrality intervals more peripheral than 60\% and $\pt>3~\GeV$,
the template-fit \vn\ have smaller values for the even-order harmonics
and larger values for the odd-order harmonics, compared to the \tpc\ results.
This is consistent with the removal of bias from dijets, which enhances (suppresses)
even-order (odd-order) harmonics.

\FloatBarrier
 
\subsection{Scaling behavior of \vn}
\subsubsection{Scaling of \vn(\pt) across different centralities}
 
In a recent ATLAS paper~\cite{HION-2016-06} on \vn\ measurements in
\PbPb\ collisions, it was observed that the $\vn(\pt)$ at different
centralities in \PbPb\ collisions had a very similar
$\pt$ dependence.
In fact, after scaling the \pt\ and \vn\ axes by multiplicative factors that depend only on centrality,
the \vn(\pt)\ across different centralities
were consistent within $\sim$5\% over the 0.5--5~\GeV\ \pt\ range.
In this section, a similar study is presented for the \vn\ measured in \XeXe\ collisions.
 
\begin{figure}[H]
\centering
\includegraphics[width=0.90\linewidth]{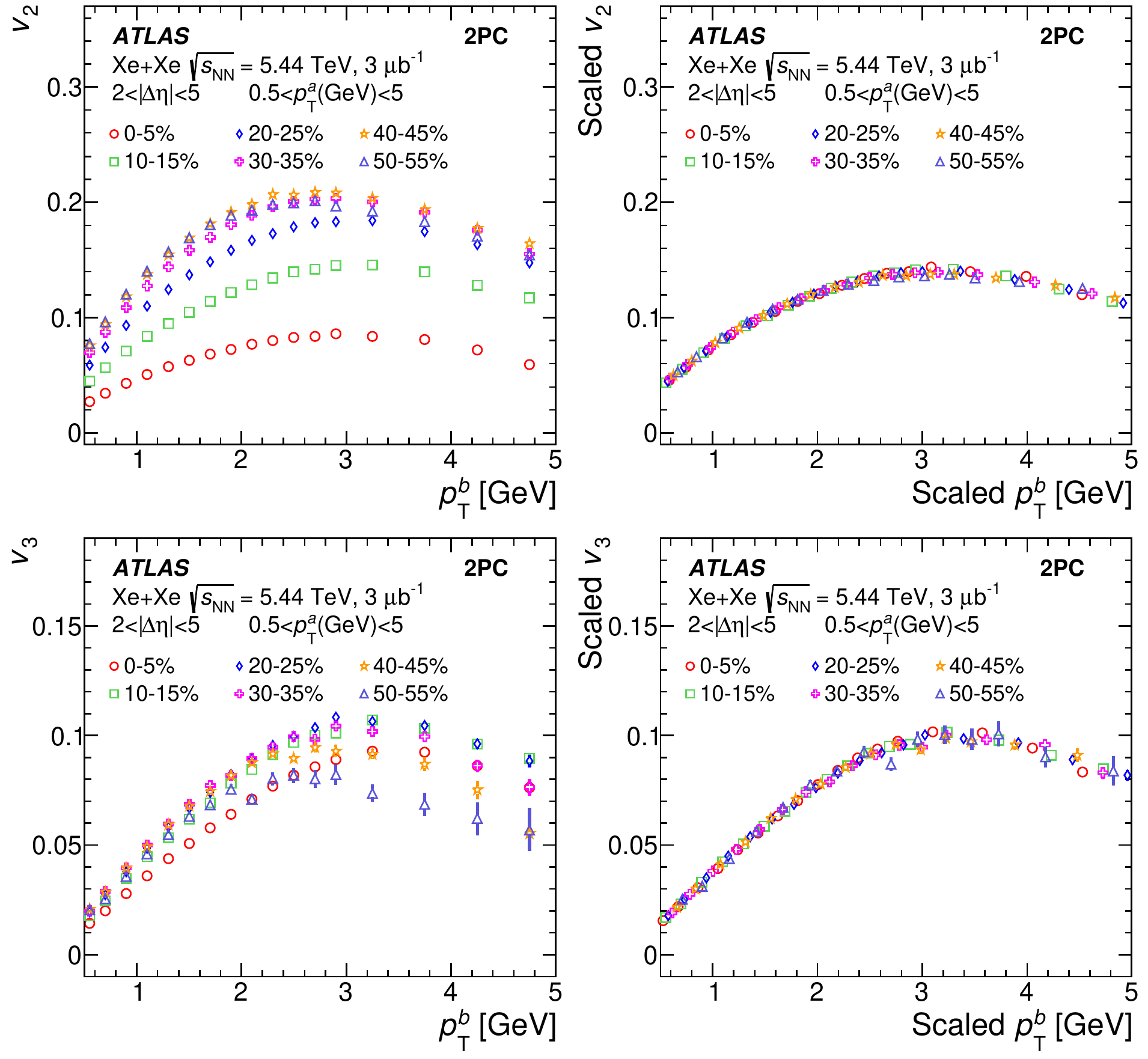}
\caption{
Left panels : The $v_{n}(p_{\mathrm {T}})$ for several centrality intervals.
Right panels: the scaled $v_{n}(p_{\mathrm {T}})$ for the same centrality intervals.
The top and bottom rows correspond to $n=2$ and 3, respectively.
The \vn\ values are obtained with the 2PC method.
The error bars indicate the quadrature sum of statistical and systematic uncertainties.
}
\label{fig:v2_scaled-1}
\end{figure}

The left panels of Figure~\ref{fig:v2_scaled-1} show the $\vn(\pt)$
for several centrality intervals and for $n=2$ and 3 obtained via the \tpc\ method.
The $v_{n}(\pt)$ are then scaled along the $x$- and $y$-axes to match
their shapes across the different centrality intervals.
For this matching, the 0--60\% centrality interval is chosen as the
reference and the $\vn(\pt)$ for the individual centralities are scaled
to best match the $\vn(\pt)$ in the 0--60\% centrality interval over the
0.5--5~\GeV $\pt$ range.
The fitting is done over the 0.5--5~\GeV\ $\pt$ range, with the $\chi^{2}$
minimization taking into account the combined statistical and systematic uncertainties,
and treating the scale factors along the \pt\ and \vn\ axes as the fit parameters.
The right panels of Figure~\ref{fig:v2_scaled-1} show the corresponding scaled \vn.
The scaled \vn\ match within $\pm$4\% across most of the shown \pt\ range.

\begin{figure}[H]
\centering
\includegraphics[width=0.95\linewidth]{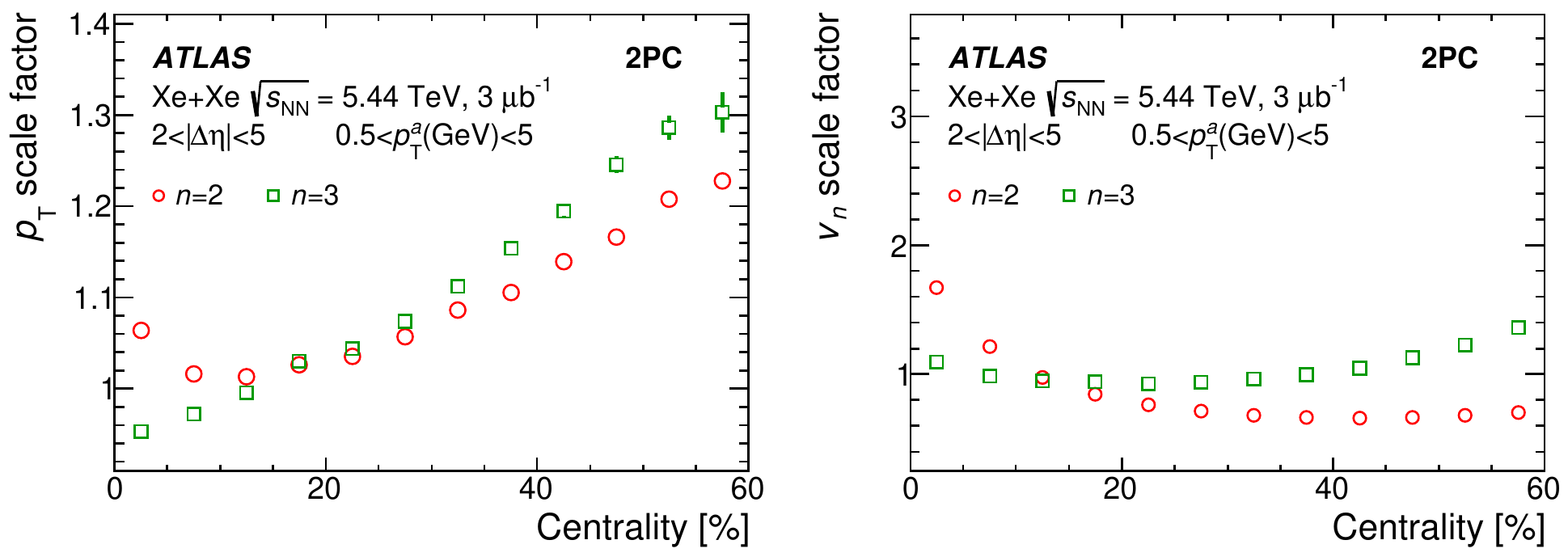}
\caption{
Left panel: the $p_{\mathrm {T}}$ scale factors for the $v_{2}(p_{\mathrm {T}})$ and $v_{3}(p_{\mathrm {T}})$
(see text) as a function of the collision centrality.
Right panel: the $v_n$ scale factors.
The error bars on the scale factors indicate fit uncertainties.
}
\label{fig:scale_factors}
\end{figure}

Figure~\ref{fig:scale_factors} shows the $\pt$ and \vn\ scale
factors obtained for $v_{2}$ and $v_{3}$ as a function of centrality.
The $\pt$ scale factors increase from central to peripheral events
over the 10--60\% centrality range.
Over this interval the values of the scale factors for the two
harmonics match within $\pm$8\%.
However, in the most central events (0--10\%) some significant deviation is
observed between the two scale factors, where the scale factor for the
\vtwo\ increases for more-central events, while that for
the \vthree\ decreases.
This deviation may be due to larger jet-bias effects in the $v_{2}$
as compared to $v_{3}$, but requires additional investigation to properly understand its origin.
The \vn\ scale factors are quite different between the two harmonics,
and show a much larger variation for \vtwo\ compared to \vthree.
This scale factor is mostly affected by the changing collision geometry;
the ellipticity of the collision geometry changes considerably with
the collision centrality leading to significantly larger variations
in the scale factor for \vtwo\ compared to that for \vthree.
 
\begin{figure}[H]
\centering
\includegraphics[width=0.95\linewidth]{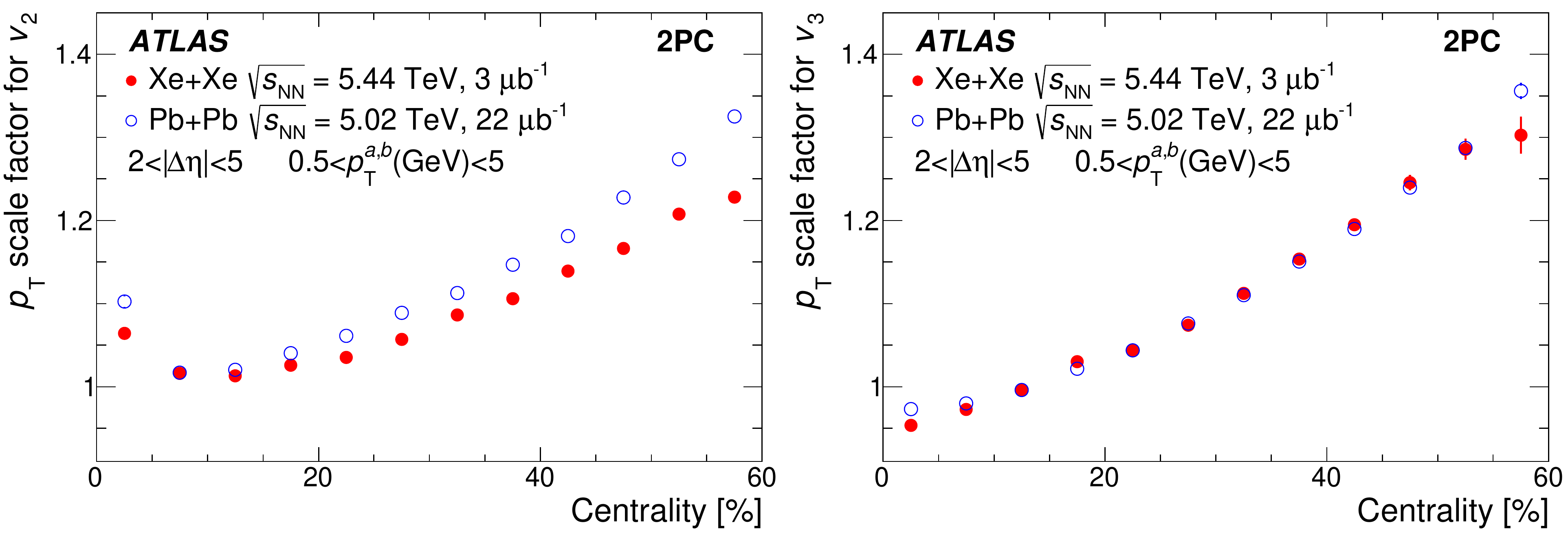}
\caption{
Comparison of the $p_{\mathrm {T}}$ scale factors for the Xe+Xe and Pb+Pb data
as a function of centrality.
The left panel shows the comparisons for $v_{2}$ and the right panel
shows the comparisons for $v_{3}$.
The error bars on the scale factors indicate fit uncertainties.
}
\label{fig:scale_factors_pbpb_xexe}
\end{figure}

Finally, Figure~\ref{fig:scale_factors_pbpb_xexe} compares the \pt\ scale
factors obtained in \XeXe\ collisions with those obtained
in \PbPb\ collisions from Ref.~\cite{HION-2016-06}, as a function of centrality.
The scale factors follow a very similar trend as a function of centrality.
In fact, for \vthree\ the values of the \pt\ scale factor are typically consistent within $\sim$2\%.
The comparison of this scaling and the corresponding scale factors
for \XeXe\ and \PbPb\ collisions should be able to provide
additional constraints on theoretical models.

\subsubsection{Scaling of \vn(\pt) with the harmonic order $n$}
\begin{figure}
\includegraphics[width=1.0\linewidth]{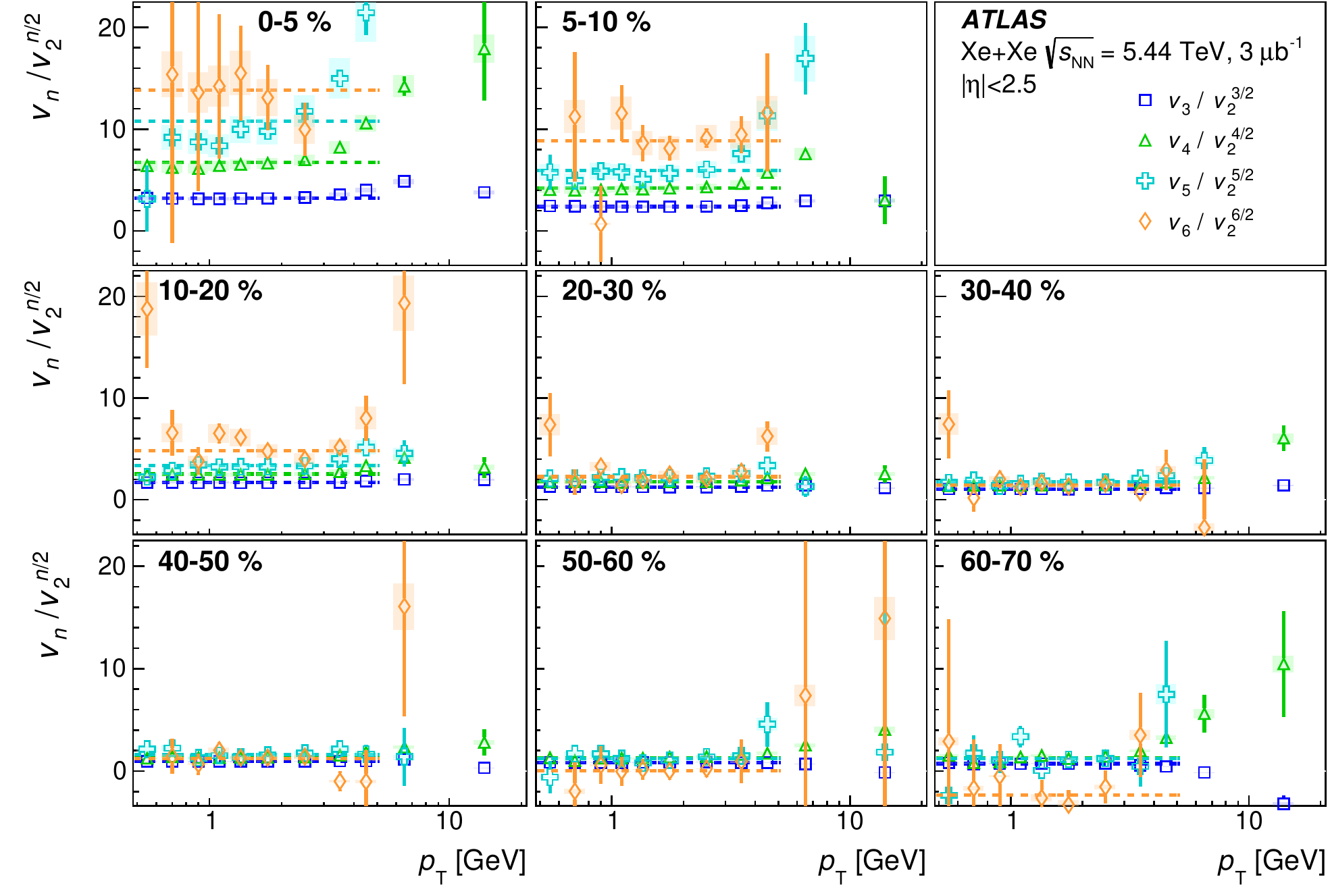}
\caption{
The $p_{\mathrm {T}}$ dependence of scaled flow harmonics, $v_{n}/v_{2}^{n/2}$, as a function
of centrality in Xe+Xe collisions.
The error bars denote statistical errors and the shaded bands
represent the systematic uncertainty.
The horizontal dashed lines represent fits to a constant,
over the 0.6--5~\GeV\ $p_{\mathrm {T}}$ range.
The $v_n$ values are measured with the SP method.
}
\label{vn_v2_pt_all}
\end{figure}

\begin{figure}
\includegraphics[width=1.0\linewidth]{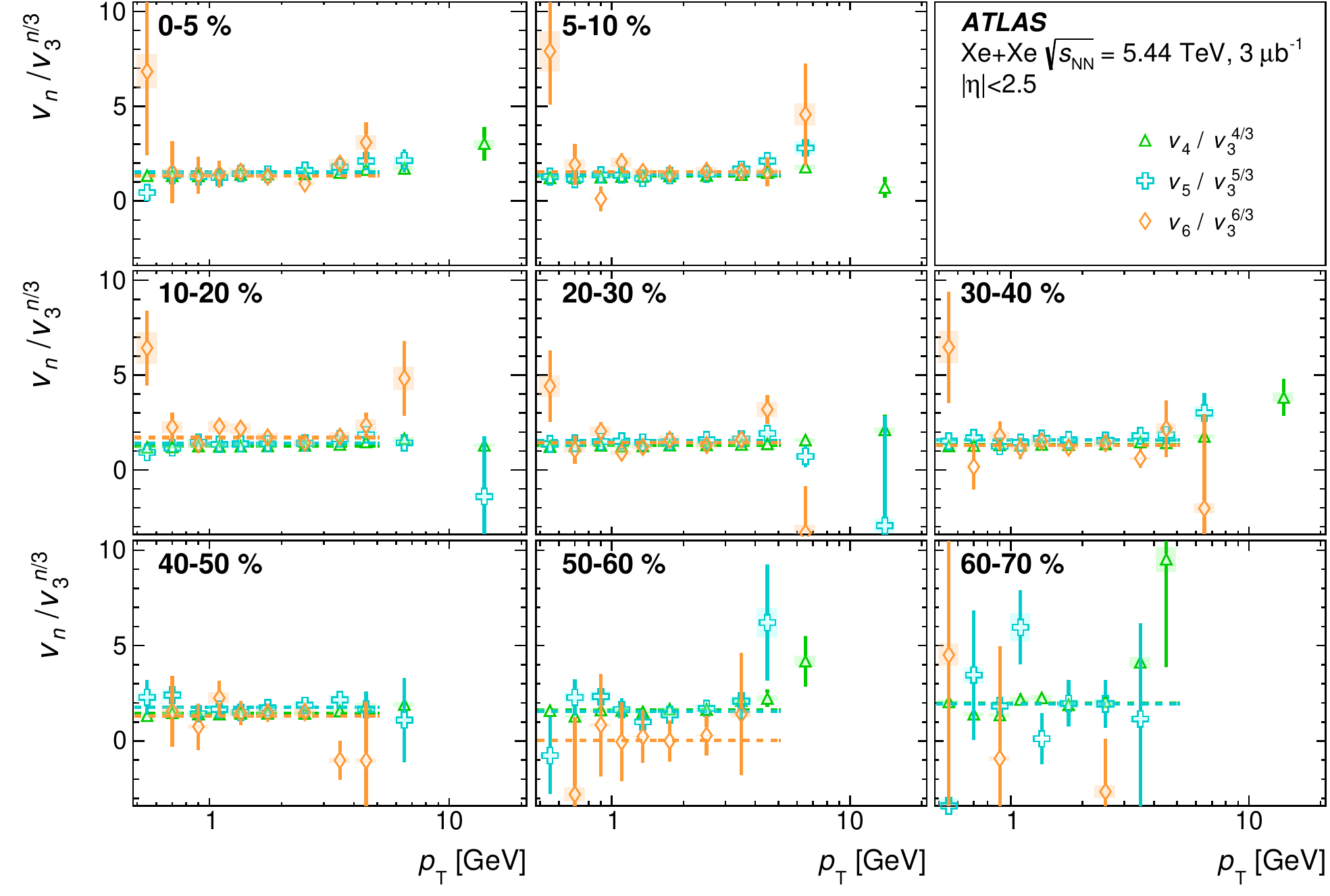}
\caption{
The $p_{\mathrm {T}}$ dependence of scaled flow harmonics, $v_{n}/v_{3}^{n/3}$, as a function
of centrality in Xe+Xe collisions.
The error bars denote statistical errors and the shaded bands
represent the systematic uncertainty.
The horizontal dashed lines represent fits to a constant,
over the 0.6--5~\GeV\ $p_{\mathrm{T}}$ range.
The $v_n$ values are measured with the SP method.
}
\label{vn_v3_pt_all}
\end{figure}

\begin{figure}
\includegraphics[width=1.0\linewidth]{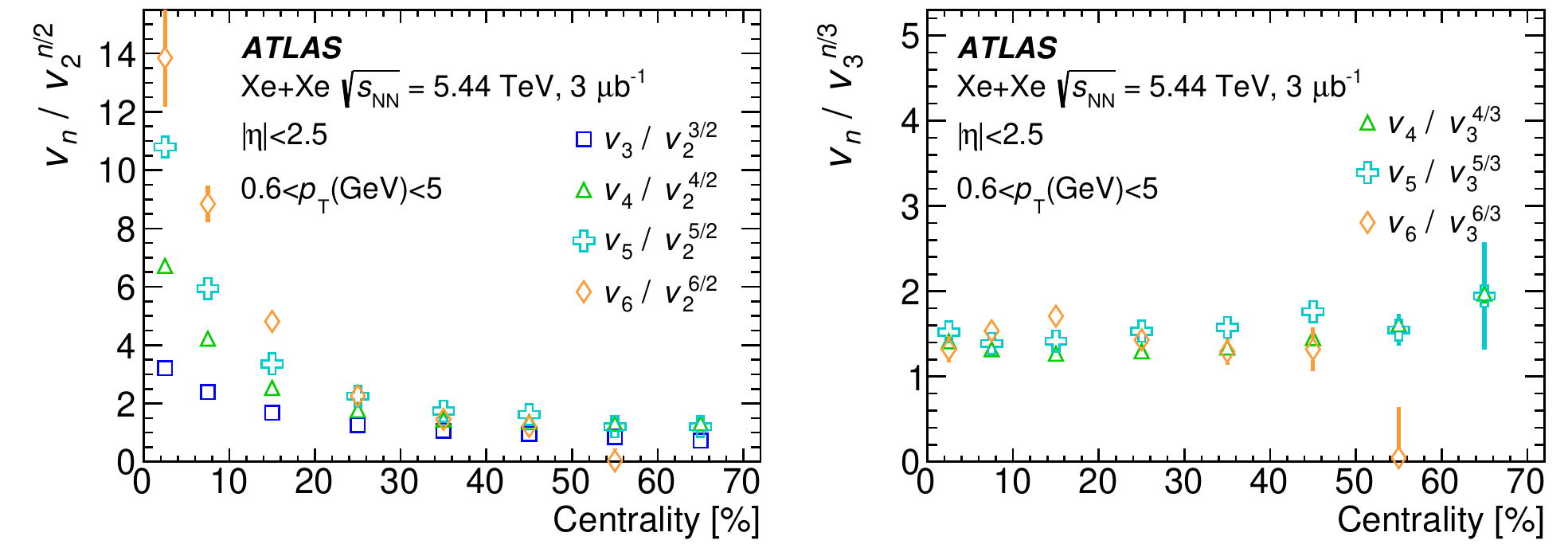}
\caption{
The centrality dependence of mean values of scaled harmonics (left) $v_{n}/v_{2}^{n/2}$
and (right) $v_{n}/v_{3}^{n/3}$ calculated in the 0.6--5~\GeV\ $p_{\mathrm{T}}$ range.
The errors represent the uncertainty of the fit and do not account for
systematic uncertainties.
}
\label{vn_ratio_v2_v3_cent}
\end{figure}

Another scaling previously observed in the \vn\ measurements in Au+Au
and \PbPb\ collisions is that the \vn\ as a function of \pt\
qualitatively follow the power-law relation $v_{n}/(v_{m})^{n/m}=\mathrm{constant}$~\cite{Adams:2003zg,
Adare:2010ux,HION-2011-01,Acharya:2018lmh}.
This scaling is demonstrated in Figure~\ref{vn_v2_pt_all} for the \vn\ in \XeXe\ collisions,
for $n=\textrm{3--6}$ and $m=2$, using the \vn\ values obtained with the SP method.
It is observed that at least in the 0.6--5~\GeV\ \pt\ range
the ratios do not depend on the transverse momentum.
The scaling often holds to considerably higher \pt\ values,
depending on the harmonic order and the centrality interval.
The mean values of $v_{n}/v_{2}^{n/2}$ are large in the most central events
and become much smaller, even less than one, in the peripheral events.
There is also a clear ordering of the ratios over the 0--30\% centrality range,
with $v_{6}/v_{2}^{6/2}$ the largest and $v_{3}/v_{2}^{3/2}$ the smallest.
The lack of \pT dependence of harmonics scaled by $v_{2}$ suggests
testing for scaling with $v_{3}$.
In Figure~\ref{vn_v3_pt_all} the flow harmonics scaled by triangular flow,
$v_{n}/v_{3}^{n/3}$, are shown.
In this case, not only are the scaled values constant as a function of \pT,
but they also have similar values for different $n$.
In addition the mean values of $v_{n}/v_{3}^{n/3}$ are very similar for all centralities.
These observations are summarized in Figure~\ref{vn_ratio_v2_v3_cent}, where
the mean values of $v_{n}/v_{2}^{n/2}$ and $v_{n}/v_{3}^{n/3}$ are shown
as a function of centrality.
The harmonics scaled by elliptic flow have a considerable variation when going from central to peripheral events.
This is because the elliptic flow has considerable centrality dependence,
while the other harmonics vary slowly with centrality.
Thus the other harmonics scaled by the elliptic flow also demonstrate
considerable centrality dependence.
In contrast, scaling by $v_{3}$ gives a $\vn/v_{3}^{n/3}$ value of
about 1.4 for all centralities and harmonics of all orders larger than $n=3$.
On the basis of the latter scaling, after measuring $v_{3}$ it is possible to
also predict $v_{n}$ for $n=4$ up to at least $n=6$.
As mentioned in Ref.~\cite{Acharya:2018lmh}, the origin of this scaling
dependence is not well understood.
Further insight from theoretical calculations into its origin
would be useful.

\FloatBarrier
 
\section{Comparison with Pb+Pb measurements at $\snn=5.02$~\TeV\label{sec:comparison}}
 
\begin{figure}
\begin{centering}
\includegraphics[width=0.95\linewidth]{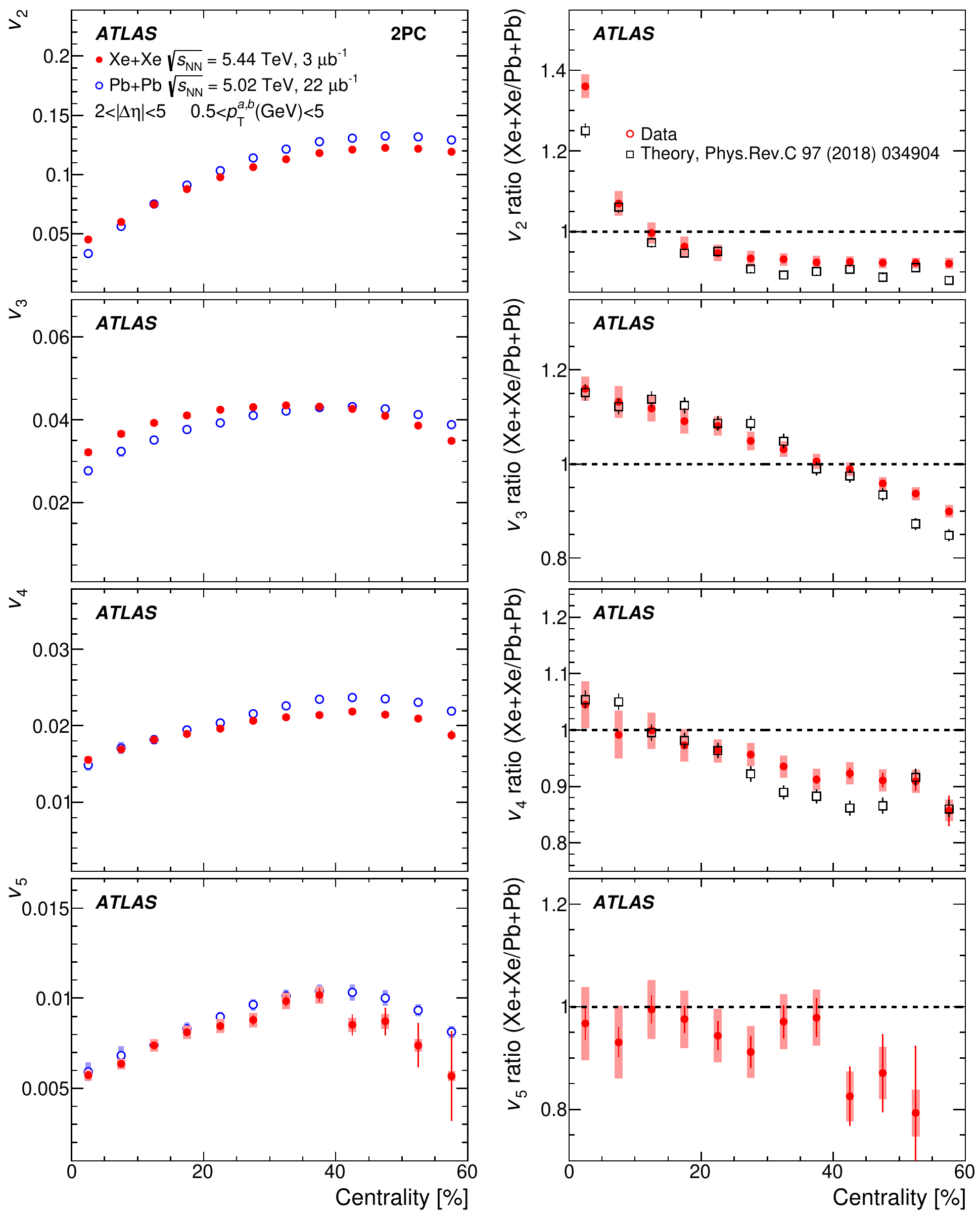}
\end{centering}
\caption{
Comparisons of the $v_n$ measured in Xe+Xe collisions with those measured in Pb+Pb collisions
at $\sqrt{s_{\mathrm{NN}}}=5.02$~\TeV.
The results are plotted as a function of centrality,
and obtained using the 2PC method.
From top to bottom, each row corresponds to a different $n$.
The right panels show the ratio of the Xe+Xe to the Pb+Pb $v_n$ values.
The plots are for $0.5~\GeV<p_{\mathrm {T}}^{\mathrm{b}}<5~\GeV$.
The error bars and bands represent statistical and systematic uncertainties, respectively.
For the ratio plots, the correlated systematic uncertainties between
the Xe+Xe and Pb+Pb results are taken into account.
}
\label{fig:xexe_pbpb_compare}
\end{figure}

\begin{figure}
\centering
\includegraphics[width=1.0\linewidth]{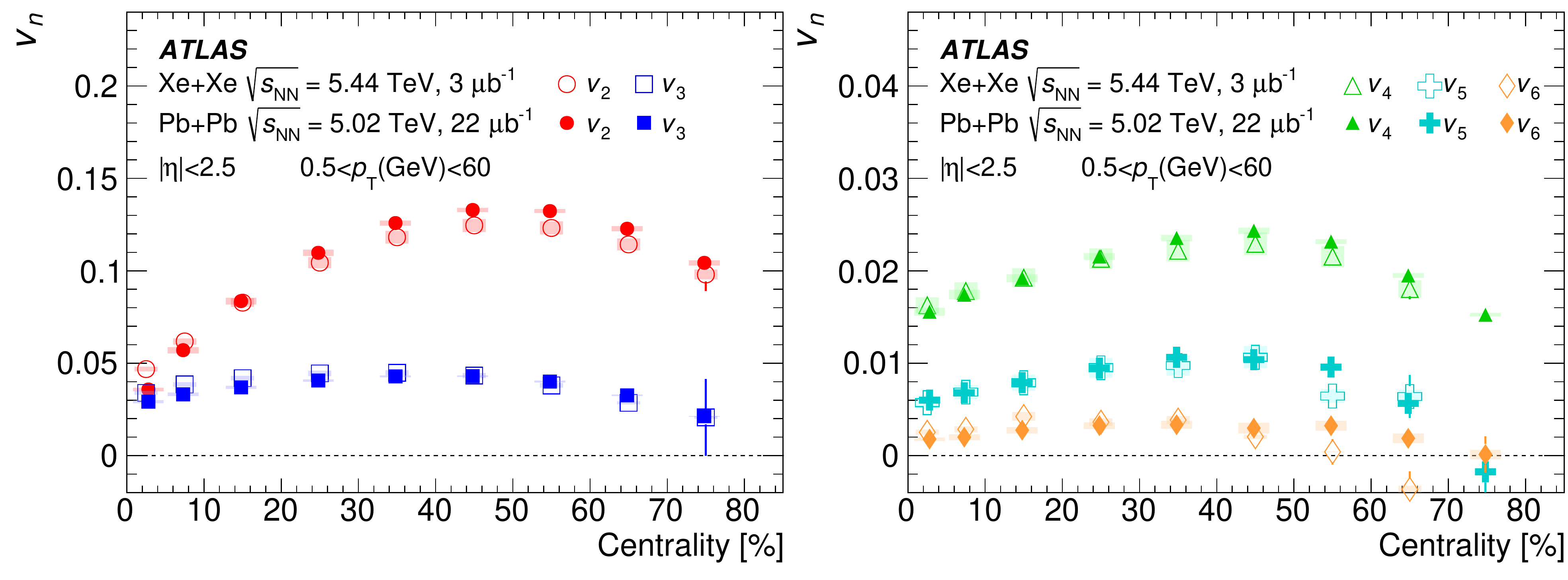}
\caption{
Comparison of the $v_{n}$ obtained with the SP method, as a function of
collision centrality, for Xe+Xe collisions at
$\sqrt{s_{\mathrm{NN}}}=5.44$~\TeV\ (open markers)
and Pb+Pb collisions at $\sqrt{s_{\mathrm{NN}}}=5.02$~\TeV\ (solid markers).
The left panel shows the comparison for $v_2$--$v_3$ and
the right panel shows the comparison for $v_4$--$v_6$.
The error bars and bands represent statistical and systematic
uncertainties, respectively.
}
\label{fig:xexe_pbpb_compare_cent_SP}
\end{figure}

Figure\,\ref{fig:xexe_pbpb_compare} shows comparisons of the \vn\ measured
in \XeXe\ collisions at $\snn=5.44$~\TeV\ with the corresponding
measurements in \PbPb\ collisions at $\snn=5.02$~\TeV, from Ref.~\cite{HION-2016-06},
as a function of centrality.
The comparisons are for measurements performed using the \tpc\ method.
For \vtwo--\vfour\ the figures also show comparisons with theoretical
calculations from Ref.\,\cite{Giacalone:2017dud}.
Figure~\ref{fig:xexe_pbpb_compare_cent_SP} shows similar comparisons for
the SP method.
In the 0--5\% central events, the \XeXe\ values for \vtwo\ are $\sim$35\% larger than
the \PbPb\ values.
This is seen most clearly from the ratio plots shown in Figure~\ref{fig:xexe_pbpb_compare}.
For less central collisions, the ratio decreases and becomes
smaller than one by the 10--15\% centrality interval.
For more-peripheral events, the ratio keeps decreasing, but at a lower
rate, and becomes roughly 0.9 by the 50--60\% centrality interval.
For \vthree\ the \XeXe\ values are larger than the \PbPb\ values over
the 0--30\% centrality interval, become comparable to the \PbPb\ values
over the 30--40\% centrality interval, and become smaller than the
\PbPb\ values for more-peripheral events.
In this case the ratio in the 0--5\% most central events is smaller
than for \vtwo, and decreases almost linearly over the 0--70\% centrality range.
For \vfour\ the \XeXe\ values are only marginally larger in the top
0--5\% central events.
The ratio for \vfour\ becomes comparable to, or less than one by the 5--10\%
centrality and continues to decrease for more-peripheral events.
For \vfive\ the \XeXe\ values are smaller throughout.
For \vtwo--\vfour, the measured ratios are quite consistent with the
theoretical predictions from Ref.~\cite{Giacalone:2017dud}, and consistent with
measurements from the ALICE and CMS Collaborations~\cite{Acharya:2018ihu,CMS-HIN-18-001}.
The observed trends can be explained as follows. Since \XeXe\ is a smaller system
than \PbPb, the impact of fluctuations is considerably more important.
The fluctuations increase the initial eccentricities of the collision
geometry and this effect contributes to enhancing the \vn\ in \XeXe\
compared to \PbPb\ at the same centrality.
Additionally the Xe nucleus may have a slight prolate
deformation~\cite{Moller:2015fba},
which would lead to larger \vtwo\ in most central \XeXe\
events, compared to \PbPb.
However, because \XeXe\ is a smaller collision system, the viscous effects,
which suppress the \vn, are larger and play a bigger role with
increasing harmonic order and for less central events.
In the most central events, the effect of the increased fluctuations
wins for \vtwo.
But with increasing harmonic order and/or for less central collisions,
eventually, the viscous effects lower the \vn\ values in \XeXe\
compared to the \vn\ values in \PbPb.
Over the centrality and \pt\ range shown in Figure~\ref{fig:xexe_pbpb_compare},
the template-fit measurements give results that are similar to those of the \tpc\ method, and are
not shown separately, for brevity.
 
\begin{figure}
\begin{centering}
\includegraphics[width=0.95\linewidth]{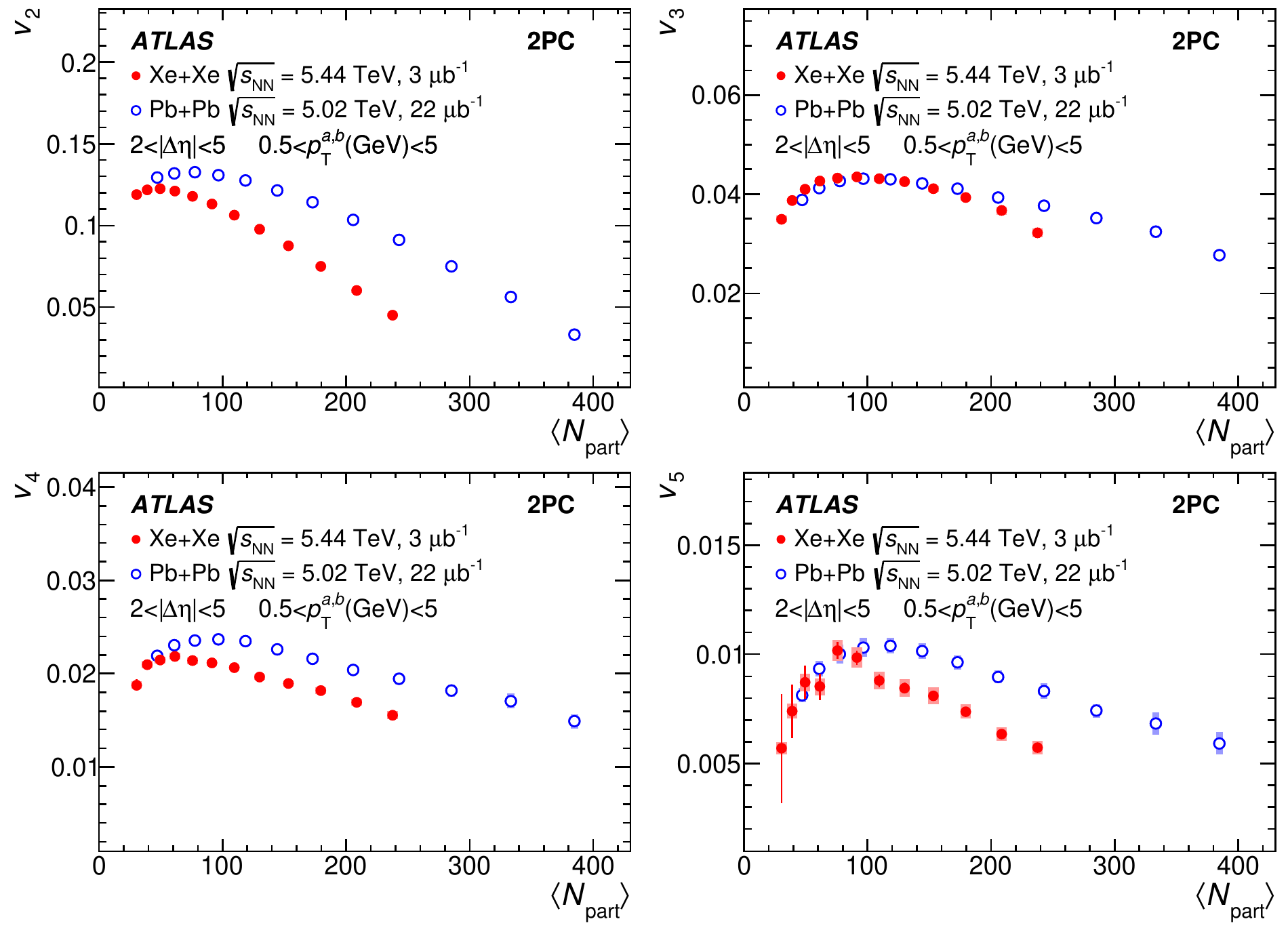}
\end{centering}
\caption{
Comparisons of the $v_n$ as a function of $\langle N_{\mathrm{part}} \rangle$
measured in Xe+Xe collisions with those measured in
Pb+Pb collisions at $\sqrt{s_{\mathrm{NN}}}=5.02$~\TeV.
Each panel corresponds to a different harmonic order $n$.
The plots are for the 2PC method and for $0.5~\GeV<p_{\mathrm {T}}^{\mathrm{b}}<5$~\GeV.
The error bars and bands represent statistical and systematic
uncertainties, respectively.
}
\label{fig:xexe_pbpb_compare_npart}
\end{figure}

\begin{figure}
\centering
\includegraphics[width=1.0\linewidth]{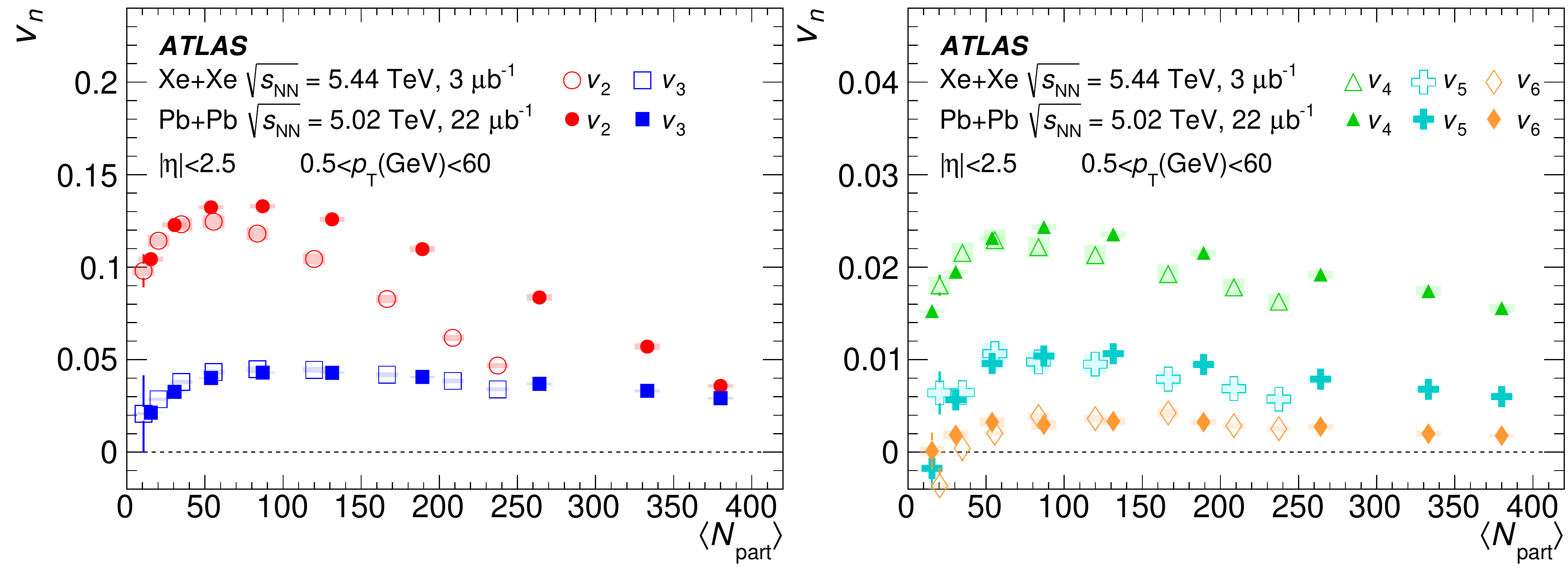}
\caption{
Comparison of the $v_{n}$ obtained with the SP method,
as a function of $N_{\mathrm{part}}$,
for Xe+Xe collisions at $\sqrt{s_{\mathrm{NN}}}=5.44$~\TeV\ (open markers) and
Pb+Pb collisions at $\sqrt{s_{\mathrm{NN}}}=5.02$~\TeV\ (solid markers).
The left panel shows the comparison for $v_2$--$v_3$ and
the right panel shows the comparison for $v_4$--$v_6$.
The error bars and bands represent statistical and systematic
uncertainties, respectively.
}
\label{fig:xexe_pbpb_compare_npart_SP}
\end{figure}

Figures~\ref{fig:xexe_pbpb_compare_npart} and ~\ref{fig:xexe_pbpb_compare_npart_SP}
show the same results but as a function of \npart.
In general, at the same \npart\ the \XeXe\ and \PbPb\ \vn\ are rather different.
One exception is the \vthree, whose values are comparable
between \XeXe\ and \PbPb\ for $\npart<200$.
The difference for the \vtwo\ arises because of the very different shapes of the
average collision geometry between \XeXe\ and \PbPb\ at the same \npart.
At the same \npart, the \PbPb\ events correspond to collisions
with significantly larger elliptic deformation.
On the other hand, the \vthree\ is driven purely by event-by-event fluctuations
in the initial geometry, and these fluctuations are expected to depend mostly on \npart,
leading to similar \vthree\ values at the same \npart.
The higher-order harmonics for $n>3$ are driven by event-by-event
fluctuations, as well as nonlinear response to lower-order
eccentricities~\cite{HION-2012-03,HION-2014-03}.
Since the second-order eccentricity for the \XeXe\ and \PbPb\ collisions is
considerably different at the same \npart, the nonlinear response
to the second-order eccentricity, which contributes to the \vfour\ and \vfive,
is different.
The \vfour\ and \vfive\ are thus not expected to be similar at the same \npart.
The difference in the \vthree\ values at the highest \XeXe\ \npart\ values
is intriguing and may be related to centrality fluctuations discussed in Refs.~\cite{HION-2017-09,Zhou:2018fxx}.
Further input from theoretical calculations to understand this feature would be useful.

\begin{figure}[H]
\begin{centering}
\includegraphics[width=0.95\linewidth]{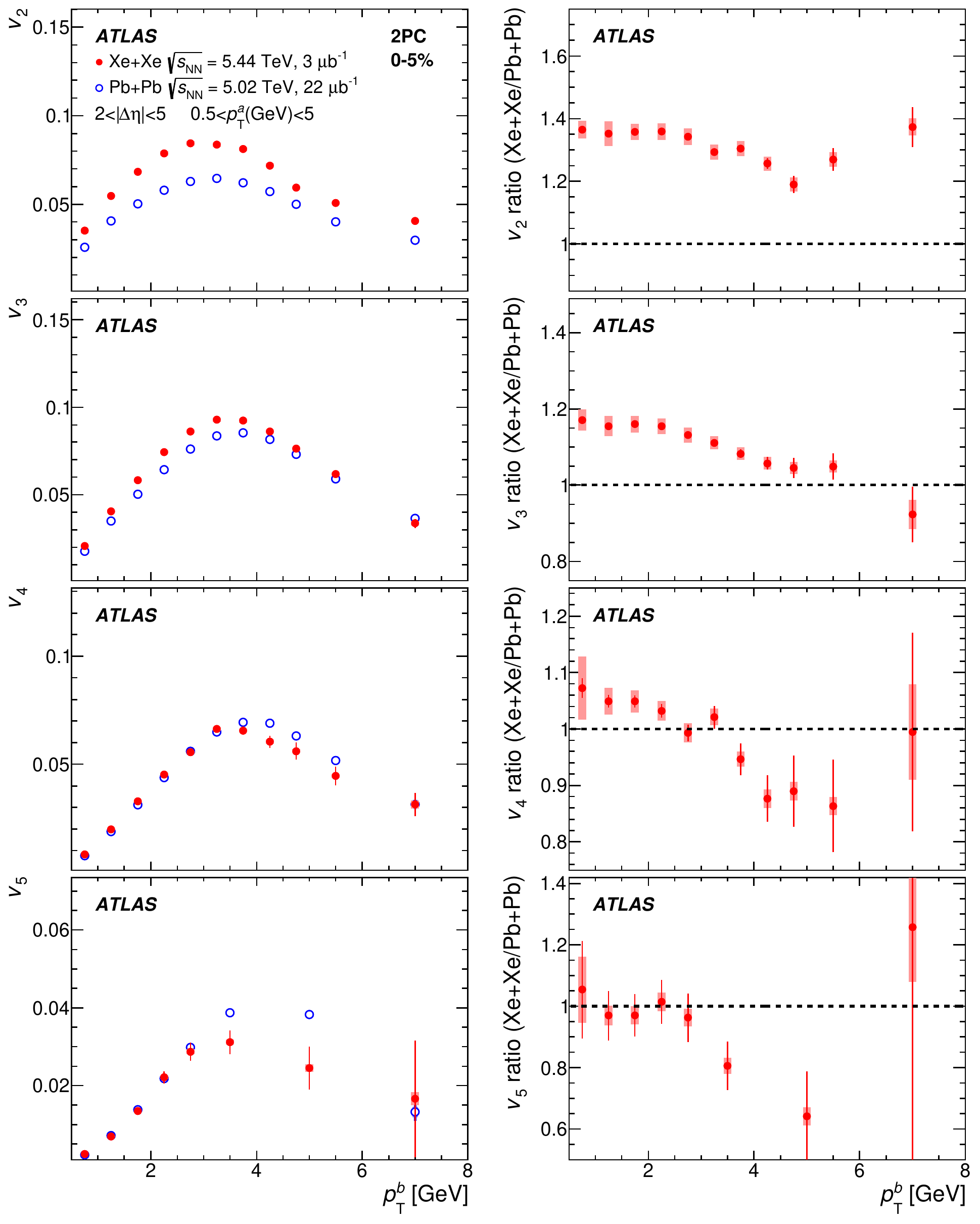}
\end{centering}
\caption{
Comparisons of the $v_{n}$ as a function of $p_{\mathrm {T}}$ between
Pb+Pb collisions at $\sqrt{s_{\mathrm{NN}}}=5.02$~\TeV\
and Xe+Xe at $\sqrt{s_{\mathrm{NN}}}=5.44$~\TeV.
The results are for the 0--5\% centrality interval and
are obtained with the 2PC method.
From top to bottom, each row corresponds to a different harmonic order $n$.
The right panels show the ratio of the Xe+Xe to the Pb+Pb $v_n$ values.
The error bars and bands represent statistical and systematic
uncertainties, respectively.
For the ratio plots, the correlated systematic uncertainties between
the Xe+Xe and Pb+Pb results are taken into account.
}
\label{fig:xexe_pbpb_compare_ptdep}
\end{figure}

Figure~\ref{fig:xexe_pbpb_compare_ptdep} shows a similar comparison but as a
function of \pt\ for the 0--5\% centrality interval.
The plots show that the ratio remains roughly constant over the
0.5--2~\GeV\ \pt\ range and then decreases for all harmonics.
This is indicative of increasing viscous effects and/or breakdown of
the hydrodynamic description beyond $\sim$2~\GeV, which lead to the lowering
of the ratio.

\clearpage
\clearpage

\begin{figure}[H]
\centering
\includegraphics[width=0.95\linewidth]{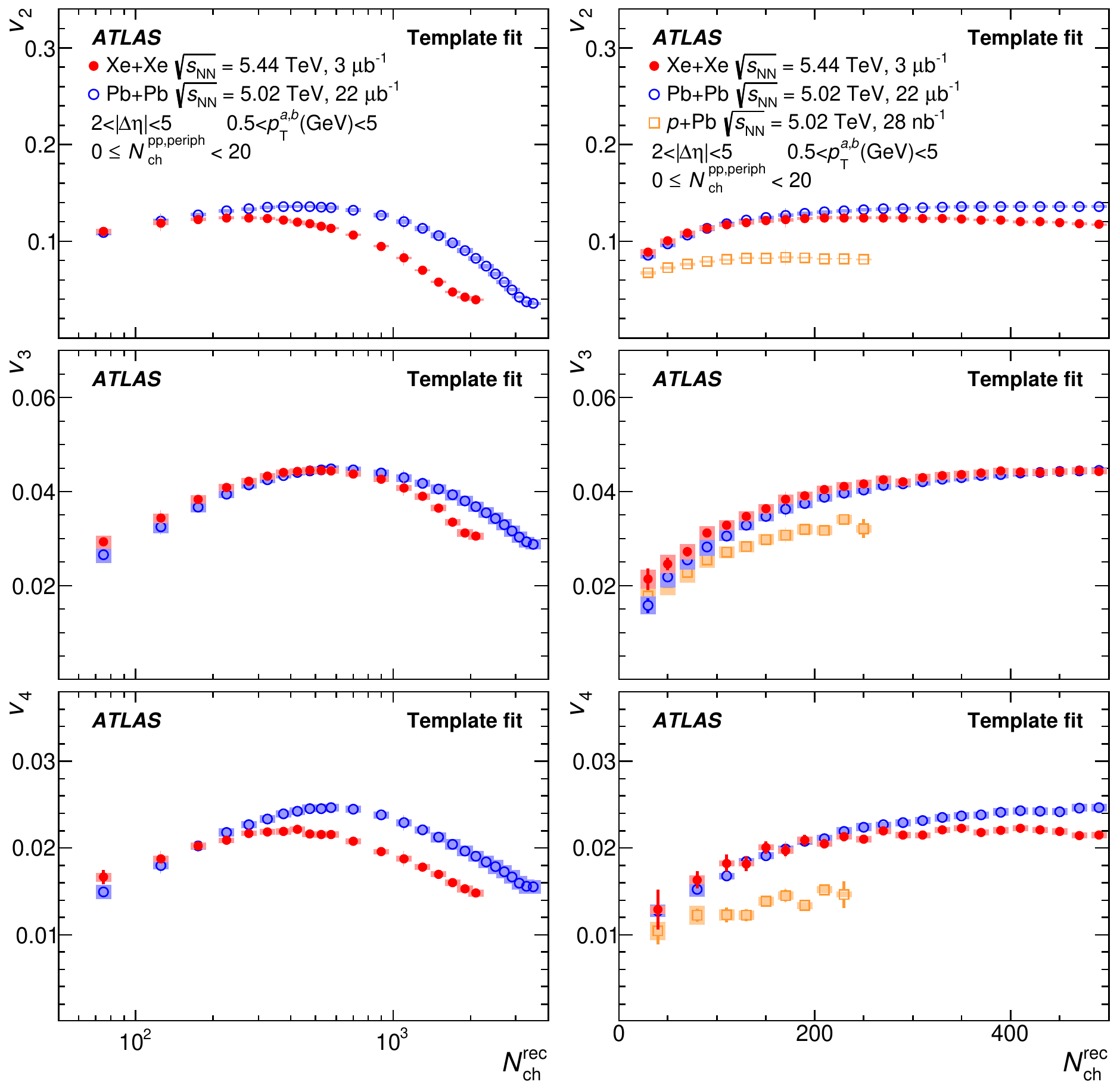}
\caption{
Comparisons of the $v_n$ measured in different collision systems as a function of multiplicity.
The left panel compares the $v_n$ obtained with the template-fit method
for the Xe+Xe measurements with those obtained in Pb+Pb collisions at $\sqrt{s_{\mathrm{NN}}} = 5.02~\TeV$.
The right panel includes comparisons with measurements in $p$+Pb at $\sqrt{s_{\mathrm{NN}}}=5.02~\TeV$,
shown over a limited multiplicity range.
The plots are for $0.5~\GeV<p_{\mathrm {T}}^{\mathrm{a,b}}<5~\GeV$.
The error bars and bands represent statistical and systematic
uncertainties, respectively.
}
\label{fig:vn_multiplicity_template_noZYAM}
\end{figure}

The left panel of Figure~\ref{fig:vn_multiplicity_template_noZYAM} shows the comparison
of the \vn\ in \XeXe\ and \PbPb\ as a function of the event multiplicity, $\nchrec$,
defined as the number of reconstructed tracks passing the
loose track selection requirements listed in Section~\ref{sec:datasets}.
Results are shown for the template-fit method.
The right panel of Figure~\ref{fig:vn_multiplicity_template_noZYAM} shows similar
results including measurements in \pPb\ but over a limited range.
For the \PbPb\ and \pPb\ template-fit measurements, the $\cperi$ (Eq.~\eqref{eq:template}) is
constructed using \pp\ events within the $\nchrec<20$ multiplicity range, similar to the \XeXe\ case.
The trends follow what was observed in the \npart\ dependence:
the \vthree\ values are quite similar between the two systems
at the same multiplicity over a large multiplicity range (up to $\sim$800),
but the values for the other harmonics are different.
The \pPb\ \vn\ values are smaller that those for \XeXe\ throughout the
overlapping \nchrec\ range.
The $\nchrec$ dependence of the \vn\ in \pPb\ collisions is weaker than
in \XeXe\ and \PbPb\ collisions.
It is interesting to note that the differences between the values (as a fraction) are
smallest for the \vthree, possibly indicating that at similar multiplicities the
geometry fluctuations in \pPb\ collisions are similar to those in heavy-ion collisions.

\FloatBarrier
\section{Summary\label{sec:summary}}
This \papertype\ presents ATLAS measurements of azimuthal anisotropy of
charged particles in \XeXe\ collisions at $\snn=5.44$~\TeV\ produced by the LHC.
The measurements are performed using an integrated luminosity
of $3~\mu\mathrm{b}^{-1}$.
The azimuthal anisotropies, quantified by the flow harmonics \vn,
are measured using the SP, \tpc\ and template-fit
methods for $n=2$--6.
The measurements are performed over wide transverse momentum
and centrality ranges.
All harmonics show a similar $\pt$ dependence, first increasing with
$\pt$ up to a maximum around 3--4~\GeV\ and then decreasing for
higher $\pt$.
Significant values of the second-order harmonic $v_{2}$
persist up to 20~\GeV.
The elliptic flow signal ($v_2$) is strongly dependent on the event
centrality and it is largest in mid-central events (30--50\%).
The higher-order harmonics show a weaker centrality dependence, which
is consistent with an anisotropy associated with fluctuations in the
initial geometry.
 
Measurements of \vn\ in peripheral heavy-ion collisions via the \tpc\ method
are known to be biased by non-flow correlations.
In particular, the even-order (odd-order) \vn\ show a rapid increase (decrease)
in value in going from mid-central to peripheral events.
This bias increases with increasing \pt.
For peripheral collisions (60--80\% centrality), the \vnn\ obtained by the template-fit
method are shown to be less biased by dijets and obey factorization better than the \tpc-\vnn.
The template-fit \vn\ are found to be significantly different from the \tpc-\vn\
at high \pt\ and in peripheral events.
However, for events over the 0--60\% centrality range and $\pt\lesssim5~\GeV$
the template fits and the \tpc\ measurements typically yield consistent results for the \vn.
 
Two types of scaling behavior observed in prior \vn\ measurements in
\PbPb\ collisions are observed to hold in the \XeXe\ collisions as well.
The first scaling is that the $\vn$ for fixed $n$ have the same shape
as a function of \pt\
across different centralities, up to an overall normalization and scaling
along the \pt\ axis.
The second scaling is that the ratio $v_{n}/v_{m}^{n/m}$
for two harmonics of order $m$ and $n$ is found to be independent
of \pt\ in a given centrality interval.
Neither scaling is understood quantitatively.
 
The measurements are also compared with prior \vn\ measurements
in \PbPb\ and \pPb\ collisions.
Compared to \PbPb\ collisions the \XeXe\ \vtwo\ is larger
in the most central collisions.
This can be attributed to larger fluctuations in \XeXe\ and a possible
deformation in the Xe nucleus, which leads to larger eccentricity
in \XeXe\ than in \PbPb.
For mid-central and peripheral events, the \vtwo\ in \XeXe\ becomes smaller
than the \vtwo\ in \PbPb, indicating the increased role of viscous effects.
This trend with centrality is also observed for the higher-order harmonics.
The ratio of the \XeXe\ \vn\ to the \PbPb\ \vn, as a function of
centrality, is found to be qualitatively consistent with
theoretical predictions.
As a function of \npart\ (or multiplicity), the \vtwo, \vfour\ and \vfive\
in \XeXe\ and \PbPb\ are observed to be very different except at
small \npart\ (multiplicity).
On the other hand, the \vthree\ values in the two systems are consistent
as a function of \npart\ except for $\npart>200$.
The \XeXe\ \vn\ measurements together with the cross-system comparisons
offer a unique opportunity to study the interplay of fluctuations in
the collision geometry and the role of viscous effects in heavy-ion
collisions, and should help to significantly improve the current
understanding of the dynamics of heavy-ion collisions.

\section*{Acknowledgments}

We thank CERN for the very successful operation of the LHC, as well as the
support staff from our institutions without whom ATLAS could not be
operated efficiently.
 
We acknowledge the support of ANPCyT, Argentina; YerPhI, Armenia; ARC, Australia; BMWFW and FWF, Austria; ANAS, Azerbaijan; SSTC, Belarus; CNPq and FAPESP, Brazil; NSERC, NRC and CFI, Canada; CERN; CONICYT, Chile; CAS, MOST and NSFC, China; COLCIENCIAS, Colombia; MSMT CR, MPO CR and VSC CR, Czech Republic; DNRF and DNSRC, Denmark; IN2P3-CNRS and CEA-DRF/IRFU, France; SRNSFG, Georgia; BMBF, HGF and MPG, Germany; GSRT, Greece; RGC and Hong Kong SAR, China; ISF and Benoziyo Center, Israel; INFN, Italy; MEXT and JSPS, Japan; CNRST, Morocco; NWO, Netherlands; RCN, Norway; MNiSW and NCN, Poland; FCT, Portugal; MNE/IFA, Romania; MES of Russia and NRC KI, Russia Federation; JINR; MESTD, Serbia; MSSR, Slovakia; ARRS and MIZ\v{S}, Slovenia; DST/NRF, South Africa; MINECO, Spain; SRC and Wallenberg Foundation, Sweden; SERI, SNSF and Cantons of Bern and Geneva, Switzerland; MOST, Taiwan; TAEK, Turkey; STFC, United Kingdom; DOE and NSF, United States of America. In addition, individual groups and members have received support from BCKDF, CANARIE, Compute Canada and CRC, Canada; ERC, ERDF, Horizon 2020, Marie Sk{\l}odowska-Curie Actions and COST, European Union; Investissements d'Avenir Labex, Investissements d'Avenir Idex and ANR, France; DFG and AvH Foundation, Germany; Herakleitos, Thales and Aristeia programmes co-financed by EU-ESF and the Greek NSRF, Greece; BSF-NSF and GIF, Israel; CERCA Programme Generalitat de Catalunya and PROMETEO Programme Generalitat Valenciana, Spain; G\"{o}ran Gustafssons Stiftelse, Sweden; The Royal Society and Leverhulme Trust, United Kingdom.
 
The crucial computing support from all WLCG partners is acknowledged gratefully, in particular from CERN, the ATLAS Tier-1 facilities at TRIUMF (Canada), NDGF (Denmark, Norway, Sweden), CC-IN2P3 (France), KIT/GridKA (Germany), INFN-CNAF (Italy), NL-T1 (Netherlands), PIC (Spain), ASGC (Taiwan), RAL (UK) and BNL (USA), the Tier-2 facilities worldwide and large non-WLCG resource providers. Major contributors of computing resources are listed in Ref.~\cite{ATL-GEN-PUB-2016-002}.
 

\printbibliography
 
\clearpage
 
\begin{flushleft}
{\Large The ATLAS Collaboration}

\bigskip

G.~Aad$^\textrm{\scriptsize 102}$,    
B.~Abbott$^\textrm{\scriptsize 129}$,    
D.C.~Abbott$^\textrm{\scriptsize 103}$,    
A.~Abed~Abud$^\textrm{\scriptsize 36}$,    
K.~Abeling$^\textrm{\scriptsize 53}$,    
D.K.~Abhayasinghe$^\textrm{\scriptsize 94}$,    
S.H.~Abidi$^\textrm{\scriptsize 167}$,    
O.S.~AbouZeid$^\textrm{\scriptsize 40}$,    
N.L.~Abraham$^\textrm{\scriptsize 156}$,    
H.~Abramowicz$^\textrm{\scriptsize 161}$
% ...
\bigskip
\\

% ...
$^{36}$CERN, Geneva; Switzerland.\\
% ...
$^{40}$Niels Bohr Institute, University of Copenhagen, Copenhagen; Denmark.\\
% ...
$^{53}$II. Physikalisches Institut, Georg-August-Universit\"{a}t G\"ottingen, G\"ottingen; Germany.\\
% ...
$^{94}$Department of Physics, Royal Holloway University of London, Egham; United Kingdom.\\
% ...
$^{102}$CPPM, Aix-Marseille Universit\'e, CNRS/IN2P3, Marseille; France.\\
$^{103}$Department of Physics, University of Massachusetts, Amherst MA; United States of America.\\
% ...
$^{129}$Homer L. Dodge Department of Physics and Astronomy, University of Oklahoma, Norman OK; United States of America.\\
% ...
$^{156}$Department of Physics and Astronomy, University of Sussex, Brighton; United Kingdom.\\
% ...
$^{161}$Raymond and Beverly Sackler School of Physics and Astronomy, Tel Aviv University, Tel Aviv; Israel.\\
% ...
$^{167}$Department of Physics, University of Toronto, Toronto ON; Canada.\\
% ...

\end{flushleft}


\end{document}